\documentclass[%
 reprint,
 amsmath,amssymb,
 aps, superscriptaddress,
]{revtex4-2}

\usepackage{color,soul}
\usepackage{graphicx}
\usepackage{dcolumn}
\usepackage{bm}
\usepackage{comment}
\usepackage{hyperref}

\begin{document}

\preprint{APS/123-QED}

\title{Near-optimal decoding algorithm for color codes using Population Annealing}

\author{Fernando Mart\'inez-Garc\'ia}
\affiliation{%
Instituto de Física Fundamental IFF-CSIC, Calle Serrano 113b, Madrid 28006, Spain
}
\email{f.martinez@iff.csic.es}

\author{Francisco Revson F. Pereira}
\affiliation{%
IQM Quantum Computers, Georg-Brauchle-Ring 23-25, 80992 Munich, Germany
}
\author{Pedro Parrado-Rodr\'iguez}%
\affiliation{%
IQM Quantum Computers, P. de la Castellana 200, Madrid 28046, Spain
}
 
\date{\today}

\begin{abstract}
The development and use of large-scale quantum computers relies on integrating quantum error-correcting (QEC) schemes into the quantum computing pipeline.  
A fundamental part of the QEC protocol is the decoding of the syndrome to identify a recovery operation with a high success rate. In this work, we implement a decoder that finds the recovery operation with the highest success probability by mapping the decoding problem to a spin system and using Population Annealing to estimate the free energy of the different error classes. We study the decoder performance on a 4.8.8 color code lattice under different noise models, including code capacity with bit-flip and depolarizing noise, and phenomenological noise, which considers noisy measurements, with performance reaching near-optimal thresholds. This decoding algorithm can be applied to a wide variety of stabilizer codes, including surface codes and quantum low-density parity-check (qLDPC) codes.

\end{abstract}

\maketitle

\section{Introduction \label{sec:introduction}}

The development of quantum computers has experienced rapid progress in the last years.
These efforts have materialized in technological and theoretical developments in multiple platforms such as superconducting circuits~\cite{clarke2008superconducting, paraoanu2014qsimSuperconducting, wendin2017quantum, kjaergaard2020superconducting}, ion traps~\cite{cirac1995quantum, haffner2008quantum, barreiro-nature-470-486, lanyon2011digitalQSim, brown2016co, bruzewicz2019trapped}, neutral atoms~\cite{jaksch2000fast, saffman2010quantum, anderson2011trapping, bluvstein2022quantum, bluvstein2024logical} or photonic devices~\cite{barz2015quantum}. While these developments continue to improve the quality of quantum processors, the main challenge towards practical quantum advantage lies in the fragile nature of quantum states. The realization of reliable, universal large-scale quantum computations thus requires the development of fault-tolerant quantum error correction (QEC) protocols capable of detecting and correcting the effects of noise, and unlocking the most powerful capabilities of quantum computation~\cite{nielsen-book,almudever2017challenges,ratcliff2017quantumchallenges, Terhal2015}. Stabilizer codes~\cite{gottesman1997stabilizer} stand as one of the most powerful tools for QEC. These QEC codes encode logical information in non-local degrees of freedom of multi-qubit states, and allow the detection and correction of errors through the measurement of stabilizer operators (or parity checks). The set of measurement results for the stabilizer operators of a given QEC protocol is called  \textit{syndrome}, and the algorithm that infers a correction from the syndrome is called  \textit{decoder}. 
Currently, the most prominent families of codes include surface codes~\cite{Kitaev1997, Kitaev2003, dennis2002topological}, color codes~\cite{colorCodes2006bombin,bombin2007colorcodes3d} and qLDPC codes~\cite{gottesman2014faulttolerant,kovalev2013qLDPC,breuckmann2021qLDPC,viszlai2023qLDPC,koukoulekidis2024qldpc,bravyi2023qLDPC}. 
The threshold theorem~\cite{Aharonov2008, Shor1996, Preskill1998} states that the logical error rate can be arbitrarily suppressed with fault-tolerant protocols by scaling the size of the code, as long as the physical error rate remains under a critical threshold.

The error threshold of a given protocol has an upper bound given by the properties of the code. This \textit{optimal threshold} can be obtained by mapping the errors on the QEC protocol into a classical statistical-mechanical model~\cite{dennis2002topological, katzgraber2009error, bombin2012strong,katzgraber2011tricolored}. However, in a practical implementation, the threshold of a protocol is limited by our capacity to \textit{decode} the information from the syndrome to infer a recovery operation. 
The problem of decoding requires a classical algorithm that is fast enough to match the rate at which the syndrome is generated, avoiding a backlog problem~\cite{Terhal2015}. In addition, achieving a high threshold necessitates an algorithm with high accuracy. Therefore, there is a trade-off between decoding quality and decoding time, where more accuracy can usually be obtained at the expense of a higher computational cost.  
The study of this trade-off has encouraged the development of multiple decoding algorithms in  recent years~\cite{sarvepalli2012rescaling,wang2009graphical,stephens2014efficient,Maskara2019networks,Delfosse_2014projection,delfosse2017almostlinear,Delfosse_2020peeling,Kubica2019cellular,kubica2019restriction,Baireuther_2019, Chamberland_2018,Davaasuren_2020, chubb2021general, parrado2022rescaling,iolius2023decoding, berent2023decoding}. 

Recently, Simulated Annealing (SA)~\cite{kirkpatrick1983optimization} has been tested for the implementation of decoders for the color code~\cite{takada2024ising} and surface code~\cite{takeuchi2023comparative}. This approach is used to select a correction by finding the minimum-weight error chain compatible with the error syndrome observed. However, it is known from the mapping of QEC codes to spin systems that the optimal decoding process, also known as \textit{maximum-likelihood decoding}, can be achieved by estimating free energy values~\cite{dennis2002topological, katzgraber2009error,ohzeki2009locations,Ohzeki_2009, bombin2012strong}.

In this work, we implement a modified version of the SA algorithm, known as Population Annealing (PA)~\cite{hukushima2003population, machta2010population}, for the decoding problem. In PA, a resampling step is introduced that helps to avoid local minima and, most importantly, allows for the estimation of the free energies. Therefore, our proposed decoder can be used to find the recovery operation with the maximum success probability, pushing the threshold of the decoder close to the optimal theoretical thresholds.  We test the decoder on a triangular color code with the square-octagon (4.8.8) lattice, reaching a threshold of 10.81\% for code capacity noise (errors happen before an ideal round of stabilizer readout) with bit-flip noise, close to the estimated optimal threshold of 10.9\% found in Refs.~\cite{katzgraber2009error,Ohzeki_2009}. It surpasses the threshold of previous decoders, such as the threshold of 10.36\% obtained recently using SA~\cite{takada2024ising}, the 10.2\% obtained using the more efficient restriction decoder with MWPM~\cite{kubica2019restriction} or the 9.8\% obtained when implementing the restriction decoder using the more scalable union-find algorithm~\cite{kubica2019restriction}, which scales almost-linealy with the number of qubits used for encoding.  When applied to the case of code capacity with depolarizing noise, we achieve a threshold of 18.75\%. This result is close to the estimated optimal threshold of 18.78\%~\cite{bombin2012strong} and improves over the previous result of 18.47\% obtained using SA~\cite{takada2024ising} or the 17.5\% obtained using neural networks~\cite{Maskara2019networks}. Finally, for the phenomenological noise model, which includes errors in the stabilizer measurements, we achieve a threshold of 3.47\%, improving over the 2.9\% obtained by SA~\cite{takada2024ising} and the $2.08\%$ using the more efficient graph matching decoder~\cite{stephens2014efficient}. While higher than the $3.3\%$ optimal threshold obtained for the surface code under phenomenological noise~\cite{ohno2004phase}, the optimal threshold under this noise model was estimated at 4.8\% for the hexagonal color code lattice~\cite{katzgraber2011tricolored,andrist2016error}, far from the value  we obtain.

The manuscript is structured as follows. In Sec.~\ref{sec:background}, we introduce basic concepts related to stabilizer codes, color codes, their mapping to spin systems, and how optimal decoding can be achieved by estimating the free energy associated with a syndrome. In Sec.~\ref{sec:population_annealing}, we explain the population annealing algorithm and how it can be used to estimate free energies. In Sec.~\ref{sec:numerical_results}, we explain details of our numerical simulations and present the results of our simulations for different  error models, namely bit-flip, depolarizing, and phenomenological noise. Then, in Sec.~\ref{sec:resource_optimization}, we provide insight on how the quality of the decoding behaves with the amount of computational resources used for decoding. We also explain how these results can be used as a guide for finding appropriate values of the hyperparameters that minimize the decoding time. Finally, in Sec.~\ref{sec:conclusions} we conclude with final remarks and ideas for further extensions of this work.

\section{Background}
\label{sec:background}
\subsection{Color codes}
Stabilizer codes are a family of quantum error correcting (QEC) codes characterized by sets of operators called stabilizers. The stabilizer operators subdivide the Hilbert space of the multiqubit system into orthogonal subspaces. The logical information is encoded into one of these subspaces, called the code space. The code space is usually chosen as the subspace for which all stabilizer operators simultaneously have a $+1$ eigenvalue. Pauli errors on individual qubits anticommute with the stabilizers, and bring the state out of the code space. By measuring stabilizer operators, it is possible to detect, identify and correct errors~\cite{gottesman1997stabilizer}. The result of the stabilizer measurements is called  syndrome.

\begin{figure}[ht]
\centering
\includegraphics[width=0.95\columnwidth]{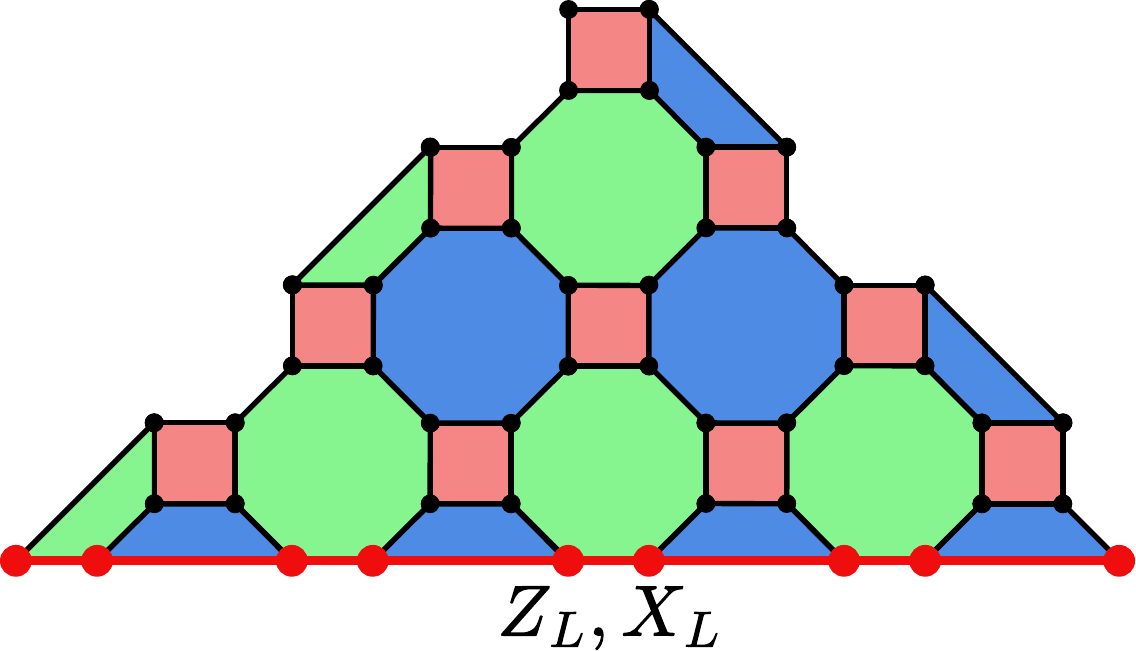}
\caption{A triangular color code of distance $d=9$, with the square-octagon lattice (4.8.8). Qubits are placed on the vertices of the lattice, and stabilizers $S_Z$ and $S_X$ have support on the vertices of each colored face. The logical operators apply to the qubits along any of the sides of the triangular lattice. An example of the support of the logical operators $X_L$ and $Z_L$ is shown in red.}
\label{fig:color_code}
\end{figure}

2D Color codes (from now on referred as color codes) are a family of stabilizer codes that can be defined on planar three-colorable lattices~\cite{colorCodes2006bombin}. Each vertex has three incident edges (except for the vertices on the corners of the lattice), and each face can be colored in one of three colors, in a way that any two faces sharing an edge have different colors (see Fig.~\ref{fig:color_code}). The vertices of the lattice represent data qubits. The colored faces represent stabilizer operators $S_Z = \prod_{i\in F} Z_i$ and $S_X = \prod_{i\in F} X_i$, where $X_i$ and $Z_i$ represent the Pauli operator $Z$ or $X$ applied on qubit $i$, and $F$ represents the set of qubits that belong to that face. The logical operators $X_L$ and $Z_L$ commute with all stabilizer operators and act on the encoded information. They can be written as a product of Pauli operators on individual qubits; e.g.,~$X_L = \prod_{i\in L_X}X_i$, where $L_X$ is a subset of qubits that forms the \textit{support} of the logical $X_L$ operator (see Fig.~\ref{fig:color_code}). For the triangular color code lattice, both $X_L$ and $Z_L$ can have their support over the same set of qubits. Note that the support of the logical operators is not unique, and can be modified by multiplying stabilizer operators. 

The decoding problem consists on the interpretation of the syndrome to infer a recovery operation that brings the state back to the code space and preserves the logical information encoded in the code with a high probability of success. In the following section, we explain how the decoding problem for color codes can be mapped to a spin system.

\subsection{Mapping the code to a spin system}
The first step of the mapping consists of the decomposition of potential error chains $\mathcal{E}$ compatible with a syndrome $S$. Note that, given an error chain that generates a syndrome $S$, we can find alternative error chains $\mathcal{E}'$ compatible with that syndrome by multiplying $\mathcal{E}$ with a product of stabilizer operators and logical operators. In this section we explain the mapping for an error model where bit-flips happen on the data qubits with probability $p$ before a single round of ideal stabilizer measurement. This derivation, as well as the derivation for depolarizing and phenomenological noise, can be found in Ref.~\cite{takada2024ising}. We consider $\mathcal{G}$ to be a complete set of stabilizer generators of the code. For a stabilizer code, an error $\mathcal{E}$ can be decomposed into a product of three components: the set of destabilizers $D(S)$ (or “pure error”) that corresponds to the syndrome $S$, a subset of stabilizer generators $G \in \mathcal{G}$ and a logical operator $L$~\cite{aaronson2004destabilizers}:	
\begin{equation}
    \mathcal{E} = D(S)\cdot G\cdot L.
\end{equation}
This expression can be rewritten in terms of binary variables, such that for each qubit $i$, $e_i$ represents if qubit $i$ belongs to the error configuration $\mathcal{E}$:
\begin{equation}\label{eq.error_on_qubit}
    e_i = D_i(S) \oplus \left(\bigoplus_{k\in Q_i} g_k\right) \oplus (L_i \cdot l),  
\end{equation}
where $D_i(S)$ represents the action of the destabilizer corresponding to syndrome $S$ on qubit $i$, $g_k$ is a binary variable that represents if the stabilizer generator $k$ is being applied (if $g_k\in G$ then $g_k=1$), $Q_i$ represents the set of stabilizer generators with support on qubit $i$, $L_i$ represents the support of the logical operator on qubit $i$, and $l$ is a binary variable that represents if the logical operator is being applied.

We can explore all error configurations compatible with a syndrome $S$ by changing the terms $g_k$ and $l$. Notably, changing $\mathcal{E}$ by applying a different subset of stabilizer generators $G'\in \mathcal{G}$ leads to alternative error configurations $\mathcal{E}'$ with an equivalent effect on the encoded information. Thus, when trying to find a correction for an error, we are only concerned about finding the error class $L$ to which it belongs, i.e., if the effect of an error corresponds to a logical operation on the encoded information.  Note that this mapping can be applied to any stabilizer code, as long as we can define a destabilizer operator $D(S)$ for any possible syndrome $S$.

The second step consists on mapping $\mathcal{E}$ to a spin configuration $\sigma$, where each spin $\sigma_k\in \{-1,+1\}$ corresponds to one of the stabilizer generators in $\mathcal{G}$: $\sigma_k = 1-2g_k$. In this way, inverting the sign of a spin would change the error configuration by the action of a stabilizer operator, leading to an equivalent error chain $\mathcal{E}'$. Similarly, we can rewrite the binary terms corresponding to the destabilizer and the logical operator in Eq.~\eqref{eq.error_on_qubit} as a coupling $J_i(l)\in \{-1, +1\}$, with $J_i(l) = (1-2D_i(S))(1-2L_i\cdot l)$. This change allows us to rewrite Eq.~\eqref{eq.error_on_qubit} in terms of the spin variables:
\begin{equation}\label{Eq:x_i}
    e_i = \frac{1}{2}\left( 1 - J_i(l) \prod_{k \in Q_i} \sigma_k\right).
\end{equation}
Using this expression, the total number of errors can be written as
\begin{equation}
    \sum_i^N e_i = \frac{1}{2}\left( N - \sum_i^N J_i(l) \prod_{k \in Q_i} \sigma_k \right),
\end{equation}
where $N$ is the total number of qubits. Therefore, by ignoring the constant term, we can write an Ising Hamiltonian for the system as
\begin{equation}
    H_l = - \sum_i^N J_i(l) \prod_{k \in Q_i} \sigma_k.
\end{equation}
Note that for each error class $l$ we find a Hamiltonian with different couplings $J_i(l)$ between the spins. By finding the spin configuration that minimizes the Hamiltonian for each value of $l$, we can then find the error configuration with the minimum number of errors, which is also the most probable error configuration.

A similar mapping can be derived for the depolarizing noise model, where Pauli errors occur with equal probability according to:
\begin{equation}\label{eq.depolarizing_channel}
    \mathcal{E}_d(\rho) = (1-p_d)\rho + \frac{p_d}{3}(X\rho X + Y \rho Y + Z\rho Z),
\end{equation}
where the map $\mathcal{E}_d(\rho)$ represents the effect of depolarizing noise on the state $\rho$ of a qubit. For this case, we must consider two logical operators, $X_L$ and $Z_L$. This leads to four homology classes: one for the case where no logical error happened, and one for each $X_L$, $Y_L$, and $Z_L$ case. The derivation of this mapping can be found in Refs.~\cite{takada2024ising,bombin2012strong}.

The last error model that we consider in this work is phenomenological noise, where bit-flips occur on the data qubits with probability $p$, and stabilizer measurement errors occur with probability $q$. For this error model, multiple rounds of stabilizer measurement are considered to protect against measurement errors. The details of the mapping for this problem are shown in Refs.~\cite{katzgraber2011tricolored,takada2024ising}. In this work, we study the $p=q$ case and for a code of distance $d$, with $d$ rounds of noisy stabilizer measurement. Finally, we note that the model can be generalized to cases with $p\neq q$~\cite{andrist2016error}.

\subsection{Optimal decoding}
\label{sec:free_energy}

While finding the minimum-weight error chain that accounts for the observed syndrome is a valid criterion for the selection of a correction, it is only an approximation to the optimal decoding scheme. The optimal decoding can be achieved in the following way: Let us consider the probability of an error $\mathcal{E}$ with an associated syndrome $S$:
\begin{equation}
    \label{Eq:bitflip_likelihood}
    P(\mathcal{E}|S) \propto \prod^N_{i=1} (1-p)^{1-e_i} p^{e_i}\propto \prod^N_{i=1} \left(\frac{p}{1-p}\right)^{e_i}.
\end{equation}
In the following, we consider that any error chain $\mathcal{E}$ can be defined by a spin configuration $\sigma$ and a homology class $l$, with a set of coefficients $J_i(l)$ given by the measured syndrome $S$. Using the expression in Eq.~\eqref{Eq:x_i} and performing a change of variable given by $\exp(-2\beta)\equiv p/(1-p)$, we obtain
\begin{equation}
    P(\mathcal{E}|S) = P(\sigma,l|S) \propto \exp\left[\sum^N_{i=1}\beta J_{i}(l) \prod_{k\in Q_i} \sigma_k\right].
\end{equation}
Finally, this probability can be written as
\begin{equation}
    P(\sigma,l|S) \propto \exp\left[-\beta H_l(\sigma)\right],
\end{equation}
which is proportional to a Boltzmann factor with energy $H_l(\sigma)$ at inverse temperature $\beta$.

To obtain the optimal correction, we do not need to find the exact original error, it is enough to perform a correction that belongs to the same homology class as the original error. If this is not the case, then the combination of the original error and the correction introduces a logical error. Thus, the best decoding strategy consists of estimating the most likely homology class. This can be obtained by performing the sum of the probabilities of all possible errors in each homology 
 class:
\begin{equation}
    P(l|S) = \sum_{\sigma}P(\sigma,l|S) \propto \sum_{\sigma}\exp\left[-\beta H_l(\sigma)\right]= \mathcal{Z}_l,
\end{equation}
where $\mathcal{Z}_l$ is the partition function of the Hamiltonian $H_l$ at inverse temperature $\beta$. The most likely homology class can then be obtained by evaluating the value of $\mathcal{Z}_{l=0}/\mathcal{Z}_{l=1}$, if this value is bigger (smaller) than 1, then the homology class $l=0$ ($l=1$) is the most likely. This result can be related to the free energy $F$ since
\begin{equation}
    -\beta F_{l=0} + \beta F_{l=1} = -\beta \Delta F = -\log\left(\mathcal{Z}_{l=1}/\mathcal{Z}_{l=0}\right),
\end{equation}
where $\Delta F$ is the difference in free energy between the two homology classes.
Therefore, by estimating $\Delta F$, we can find the most likely error class, maximizing the success probability of the decoding operation.

This method can also be applied to other error models, like depolarizing noise or phenomenological noise. For depolarizing noise, we consider Pauli errors on the physical qubits occurring with probability $p_d/3$ (see Eq.~\eqref{eq.depolarizing_channel}). Using the mapping described in Ref.~\cite{takada2024ising}, it can be found that the target inverse temperature for this model is given by
\begin{equation}
    \beta=-\frac{1}{4}\log\left(\frac{p_d/3}{1-p_d}\right).
\end{equation}
To find the optimal recovery operation, one must find the most likely homology class, given by the combinations of the logical operators $X_L$ and $Z_L$. Generally, for a code with $k$ independent logical operators, the optimal decoder should explore the $2^k$ different homology classes for the bit-flip noise case (or $4^k$ for the depolarizing noise case) to find the optimal correction.
\section{Population Annealing}
\label{sec:population_annealing}

Population annealing (PA)~\cite{hukushima2003population, machta2010population} is an algorithm closely related to the Simulated Annealing (SA) algorithm~\cite{kirkpatrick1983optimization}. Both are sequential Monte Carlo algorithms that start with a set of $R$ \textit{replicas} $\sigma^{(i)}$, $i=1,..., R$, which in our case of interest are spin configurations. These replicas have an energy associated with a given Hamiltonian and, when their spin values are initialized randomly, they can be considered as samples from a Boltzmann distribution at a temperature $T\rightarrow \infty$ or inverse temperature $\beta_0=1/T=0$. Given a replica $\sigma$ with energy $E$, it is possible to propose a change to it (e.g., a single spin flip) that results in the configuration $\sigma'$ with energy $E'$. To transform our replicas from thermal distribution samples with $\beta_0=0$ to samples corresponding to $\beta_1>\beta_0$, we accept or reject these changes with a probability given by the Metropolis-Hastings rule~\cite{robert1999monte}:
\begin{equation}
\label{eq:MH_rule}
    P_\text{accept}(\sigma'|\sigma) = \begin{cases}
        1 \quad &\text{if} \quad E' \leq E \\
        e^{-\beta_1(E'-E)} &\text{if}  \quad E' > E .
    \end{cases}
\end{equation}
 Eventually, after proposing enough changes, the resulting replicas will approximate the result of sampling from the Boltzmann distribution at inverse temperature $\beta_1$. This process can be iterated for increasing inverse temperature values $\beta_t$, with $t=1,...,N_T$, until reaching a target temperature $\beta_{N_T}$. These steps constitute the simulated annealing algorithm.

\begin{figure}[ht]
\centering
\includegraphics[width=0.95\columnwidth]{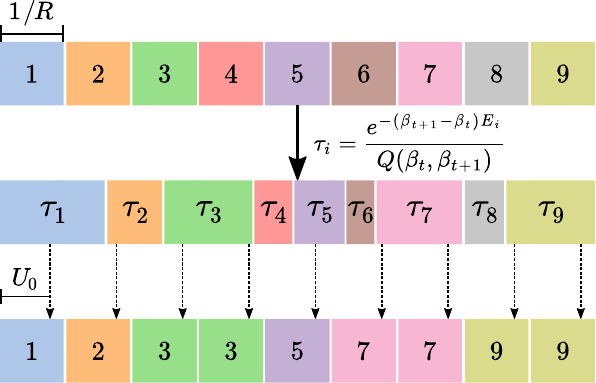}
\caption{Schematic representation of the systematic resampling procedure for a set of $R=9$ replicas with weights $\tau_i$, $i=1,...,9$. The original replicas (top row) are assigned a weight relative to their Boltzmann factor (middle row). Generating a random number $U_0\in[0,1/R)$, and using these weights and the values $U_k=U_0+k/R$, with $k=0,..., R-1$, one can obtain a new set of resampled replicas (bottom row).}
\label{fig:systematic_resampling}
\end{figure}

In the PA algorithm, when changing the temperature of the system from $\beta_t$ to $\beta_{t+1}$, a resampling between the replicas is performed by associating a probability proportional to the relative Boltzmann weights between each temperature to each replica:
\begin{equation}
    \tau_i = \frac{e^{-(\beta_{t+1} - \beta_t) E_i}}{Q(\beta_t, \beta_{t+1})},
\end{equation}
with the normalization factor
\begin{equation}
    Q(\beta_t, \beta_{t+1}) = \sum_{i=1}^R e^{-(\beta_{t+1} - \beta_t) E_i}.
\end{equation}
There are different ways of resampling using these probabilities. In this work, we chose to implement the so-called \textit{systematic resampling} approach, which keeps the number of replicas constant at all times, requires only one randomly generated number for each resampling step, and was found to introduce fewer statistical errors as compared to other resampling methods~\cite{gessert2023resampling}. To visualize how systematic resampling works, let us consider a line of unit length. Each replica can be positioned in this line with a length equal to its corresponding $\tau_i$ value (see Fig.~\ref{fig:systematic_resampling}). This resampling is implemented by generating a random number $U_0\in [0,1/R)$ and selecting $R$ positions given by the values $U_k=U_0+k/R$ with $k=0,..., R-1$. Each of these values will be associated with a replica. The set of replicas associated with the $R$ values $U_k$ will be the new set of resampled replicas. This completes the resampling process.

The resampling step and the acceptance-rejection defined by the Metropolis-Hastings rule are complementary to each other. On the one hand, when applying the SA algorithm alone, the replicas can get stuck in local minima and are less likely to escape as the temperature decreases. This reduces the number of effective replicas, resulting in a less efficient use of the computational resources. This effect is mitigated by resampling the replicas, which introduces a mechanism for these trapped replicas to escape. On the other hand, while the resampling step introduces correlations between replicas that are resampled to the same configuration, this is alleviated by the acceptance-rejection protocol, which uncorrelates the replicas step by step.

However, the most important consequence of introducing this resampling step is that it can be used to estimate the free energy, $\overline{F}(\beta_{N_T})$, at the target temperature $\beta_{N_T}$~\cite{machta2010population}. This can be obtained as:
\begin{equation}
    -\beta_{N_T} \overline{F}(\beta_{N_T})=\sum^{N_T-1}_{t=0} \ln{Q(\beta_t, \beta_{t+1})} + \ln{\Omega},
\end{equation}
where $\Omega$ is the total number of possible configurations. As explained in Sec.~\ref{sec:free_energy}, estimating the values of the free energies for different homology classes can be used to achieve optimal decoding in QEC codes for different error models. In the following section, we explain and present in detail our simulations of PA as a decoder for the color code.

\section{Numerical results}
\label{sec:numerical_results}

\subsection{Simulation details}

We simulate the decoding process on triangular color code lattices for different code distances $d$ and error rates $p$ around the expected value of the threshold corresponding to each error model. For these decoding simulations, we use $R=1000$ replicas, $N_T=100$ inverse temperature steps following a linear inverse temperature schedule, and $N_S=200$ sweeps over all the spin variables, where a sweep consists of going over each of the spins (stabilizers) of the code always in the same order and accepting or rejecting a change of its value based on Eq.~\eqref{eq:MH_rule}. We note that these values of the hyperparameters correspond to computational resources that are far above those required for the correct behaviour of the PA decoder. We use the additional resources to ensure that the decoder finds the optimal correction for all the cases. The optimization of the hyperparameters is further discussed in Sec.~\ref{sec:resource_optimization}. We also note that the PA decoder dedicates most of the computing time on the acceptance or rejection of spin-flip candidates. A non-parallelized version of our code using an AMD Ryzen 9 5950x 3.4GHz requires 7ns for each candidate. However, our implementation takes advantage of CPU parallelization, which considerably improves the algorithm speed, reducing the time per candidate to approximately $0.5$ns in our simulations. Additional methods to reduce computational time are discussed in Sec.~\ref{sec:conclusions}.

From the simulation results, we obtain the probability of a logical error $p_L$ for each value $p$ and $d$. With these results, we assume a critical scaling ansatz to estimate the corresponding threshold. Using this ansatz, we expect $p_L$ to behave as a linear function around the threshold, given by:
\begin{equation}
\label{eq:critical_ansatz}
    p_L = A + B d^{1/\nu}(p-p_{th}),
\end{equation}
where $A$ and $B$ are linear fit parameters, $\nu$ is the critical exponent, and $p_{th}$ is the value of the threshold. For this approximation to be accurate, we perform the fit considering only points close to the threshold. Also, we only consider high enough distances so that finite-size effects do not affect our analysis.

\subsection{Bit-flip noise}

\begin{figure}[t]
\centering
\includegraphics[width=0.99\columnwidth]{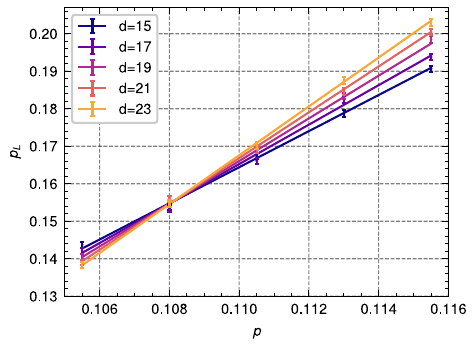}
\caption{Results from the simulation of the population annealing decoder applied to the 4.8.8 color code with bit-flip noise. The dots represent the data obtained from the simulations, with each point being obtained from $2\cdot 10^5$ decoding instances. The lines represent the corresponding linear fit described in Eq.~\eqref{eq:critical_ansatz} for different distances.} 
\label{fig:bitflip_results}
\end{figure}

As explained in Sec.~\ref{sec:free_energy}, for bit-flip noise we have to estimate the free energies associated with the two different homology classes. We simulate the decoding process under bit-flip noise to obtain the logical error probabilities for distances $d=15, 17, 19, 21, 23$. We show the results of these simulations in Fig.~\ref{fig:bitflip_results}. Fitting these results to Eq.~\eqref{eq:critical_ansatz} we obtain the following values:
\begin{equation}
\begin{split}
    A&=0.155\pm 0.002,\\
    B&=0.709 \pm 0.127,\\
    \nu&=1.41\pm 0.12,\\
    p_{th}&=0.1081 \pm 0.0003.
\end{split}
\end{equation}

For this case, we find that using the PA decoder to estimate free energies improves the quality of the decoding process as compared to finding the minimum-weight error chain using SA, for which the threshold found in Ref.~\cite{takada2024ising} is $10.359\%$. Additionally, the value that we find for the threshold is close to the optimal threshold of 10.9\% estimated in Refs.~\cite{katzgraber2009error, chubb2021general}.

\subsection{Depolarizing noise}

\begin{figure}[t]
\centering
\includegraphics[width=0.99\columnwidth]{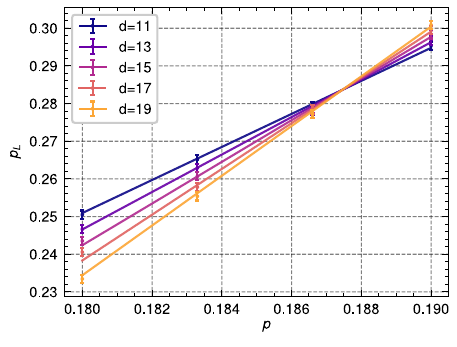}
\caption{Results from the simulation of the population annealing decoder applied to the 4.8.8 color code with depolarizing noise. The dots represent the data obtained from the simulations, with each point being obtained from $2\cdot 10^5$ decoding instances. The lines represent the corresponding linear fit described in Eq.~\eqref{eq:critical_ansatz} for different distances.}
\label{fig:depolarizing_results}
\end{figure}

For depolarizing noise, we have to estimate the free energies associated with four different homology classes. We simulate the decoding process under depolarizing noise to obtain the logical error probabilities for distances $d=11, 13, 15, 17, 19$. We show the results of these simulations in Fig.~\ref{fig:depolarizing_results}. Fitting these results to Eq.~\eqref{eq:critical_ansatz}, we obtain the following values:
\begin{equation}
\begin{split}
    A&=0.2836\pm 0.0017,\\
    B&=0.7309\pm0.1034,\\
    \nu&=1.338\pm0.092,\\
    p_{th}&=0.1875\pm 0.0003.
\end{split}
\end{equation}

For this case, we find again that using the PA decoder to estimate free energies improves the quality of the decoding process as compared to finding the minimum-weight error chain, for which the threshold found in Ref.~\cite{takada2024ising} was $18.467\%$. Additionally, the value that we find for the threshold is in close agreement with the 18.9\% optimal threshold numerically obtained for the 6.6.6 color code and the value estimated for the 4.8.8 color code of 18.78\% in Ref.~\cite{bombin2012strong}.

\subsection{Phenomenological noise}

\begin{figure}[ht]
\centering
\includegraphics[width=0.99\columnwidth]{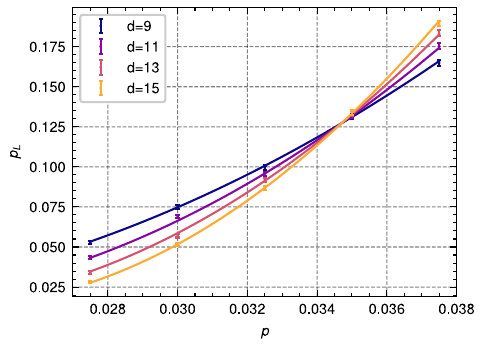}
\caption{Results from the simulation of the population annealing decoder applied to the 4.8.8 color code with phenomenological noise. The dots represent the data obtained from $5\cdot 10^4$ decoding instances. The lines represent the corresponding result of the quadratic version of the fit described in Eq.~\eqref{eq:critical_ansatz} for different distances.}
\label{fig:phenomenological_results}
\end{figure}

For phenomenological noise, we consider only bit-flip errors, so we need to estimate the free energies associated with two different homology classes. We simulate the decoding process under phenomenological noise to obtain the logical error probabilities for distances $d=9, 11, 13, 15$. We show the results of these simulations in Fig.~\ref{fig:phenomenological_results}. For the values of $p$ and $d$ used in these simulations, we observe a deviation from a linear fit. Therefore, we perform a fit similar to Eq.~\eqref{eq:critical_ansatz} but including a quadratic term, described by the parameter $C$. We obtain the following values:
\begin{equation}
\begin{split}
    A&=0.127\pm 0.002,\\
    B&=1.80\pm 0.16,\\
    C&=7.07\pm 1.13,\\
    \nu&=1.12\pm 0.04,\\
    p_{th}&=0.0347\pm 0.0002.
\end{split}
\end{equation}

Once again, we find that using the PA decoder to estimate free energies improves the quality of the decoding process as compared to finding the minimum-weight error chain, for which the threshold found in Ref.~\cite{takada2024ising} with SA was $2.90\%$. Our result sits above the optimal threshold of $3.3\%$  estimated for surface codes~\cite{ohno2004phase}. While the authors are not aware of any study of the optimal threshold for 4.8.8 color codes under phenomenological noise, it does not match the optimal threshold estimated for the hexagonal color code at 4.8\%~\cite{katzgraber2011tricolored,andrist2016error}. The reason for this discrepancy deserves further study, but it is beyond the scope of this work.


\section{Resource optimization}
\label{sec:resource_optimization}

The previous results were obtained by using more computational resources than needed to ensure that we obtained high-fidelity threshold values. However, in practice, one would want to reach a compromise between the quality of the solutions and computational resources, i.e., the time required for decoding. In the following, we study the relation between these two quantities. Although we show the analysis applied to the bit-flip noise model, the ideas presented here are directly applicable to other noise models.

\begin{figure}[ht]
\centering
\includegraphics[width=0.99\columnwidth]{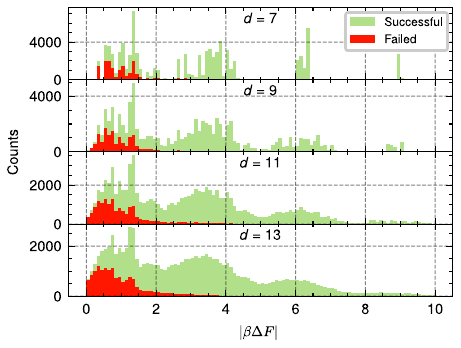}
\caption{Histogram representation used to estimate the probability densities $P_{s}(|\beta \Delta F|)$ (green) and $P_{f}(|\beta \Delta F|)$ 
 (red) for values of $d=7,9,11,13$ and $p=10.8\%$. These histograms were obtained by simulating $10^5$ instances of PA decoding for each distance. For each instance, the PA decoder used a high number of resources, specifically $R=1000$ replicas, $N_T=100$ temperature steps, and $N_S=200$ sweeps, to achieve a small error in the histograms.}
\label{fig:free_energy_histograms}
\end{figure}

The population annealing decoder for the bit-flip noise model works by estimating the free energy of the two possible homology classes and choosing the corresponding correction based on the difference between them, $\beta\Delta F$. However, since population annealing uses a limited amount of computational resources (number of replicas and spin flips), the estimate obtained has an associated variance $\text{Var}(\beta\Delta F)$. We study how this variance increases the probability of logical error in the following way: For a given bit-flip probability and code distance, we simulate several instances with a high number of resources to obtain an estimate of the distribution $P(|\beta\Delta F|)$ with sufficiently small error (see Fig.~\ref{fig:free_energy_histograms}). We know that some of these instances will correspond to successful corrections, denoted as $P_\text{s}(|\beta\Delta F|)$, while the rest will correspond to failed corrections, $P_\text{f}(|\beta\Delta F|)$. As previously explained, in a real implementation of the decoder there will be an error in the estimation of $|\beta\Delta F|$ due to using a finite number of computational resources. The value of $|\beta\Delta F|$ plus this error might result in a negative value. For these cases, the decoder finds the opposed homology class than the one which the optimal decoder would find. Therefore, an error configuration that an optimal decoder would successfully correct, can lead to a logical error with the finite-resource decoder due to the error in the estimation of $|\beta\Delta F|$. We denote the probability of this happening as $P_{\text{s}\rightarrow\text{f}}$. Similarly, the error can transform a correction that would otherwise be a failed correction into a successful one, $P_{\text{f}\rightarrow\text{s}}$. As a consequence, the logical error probability, $p_L$, is increased by $\Delta p_L$:
\begin{equation}
    p'_L = p_L + \Delta p_L,
\end{equation}
with
\begin{equation}
\label{eq:deltap_l}
    \Delta p_L= (1-p_L)P_{\text{s}\rightarrow\text{f}} - p_L P_{\text{f}\rightarrow\text{s}},
\end{equation}
where $\Delta p_L$ can be obtained by using the estimated distribution $P(|\beta\Delta F|)$ and the value of $\text{Var}(\beta\Delta F)$ associated to using a finite number of computational resources.

\begin{figure}[ht]
\centering
\includegraphics[width=0.95\columnwidth]{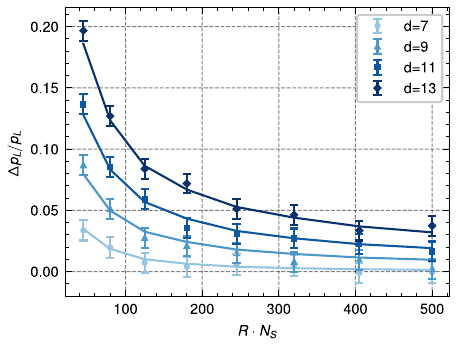}
\caption{Comparison between the values of $\Delta p_L/p_L$ estimated by using the histograms in Fig.~\ref{fig:free_energy_histograms} and Eq.~\eqref{eq:deltap_l} (points with error bars), and the corresponding values obtained by numerical simulation (continuous lines) for codes of distances $d=7, 9, 11, 13$ with $p=10.8\%$. We simulate eight values of $RN_S$ where the number of replicas $R$ and the number of sweeps $N_S$ are increased. These values are $N_S=3,4,5,6,7,8,9,10$ sweeps and $R=15,20,25,30,35,40,45,50$ replicas.}
\label{fig:error_estimation_benchmark}
\end{figure}

The previous derivation introduces a way to relate the quality of the decoding process with the error in the estimate of $\Delta F$. We have tested this relation on the color code with $d=7, 9, 11, 13$ and $p=10.8\%$. We use the PA decoder for a fixed value of $N_T=30$ temperature steps while changing the value of $R N_S$. For each of these values, we estimate the corresponding value of $\text{Var}(\beta\Delta F)$ by simulating $200$ decoding instances $100$ times each. We can use the obtained value of the variance and the histograms of $P(|\beta\Delta F|)$ to obtain the estimated value of $\Delta p_L$. We then simulate $5\cdot 10^5$ decoding instances for each value of $d$ and $R N_S$ to estimate $p'_L$ and subtract the estimated value of $p_L$ for those values of $d$ and $p$. The results obtained from our estimation method and those found from simulations are in close agreement and are shown in Fig.~\ref{fig:error_estimation_benchmark}.

\begin{figure}[ht]
\centering
\includegraphics[width=0.99\columnwidth]{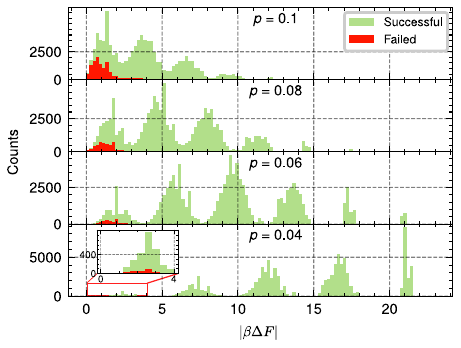}
\caption{Histogram representation used to estimate the probability densities $P_s(|\beta \Delta F|)$ and $P_f(|\beta \Delta F|)$ for a code with distance $d=11$ and different values of $p$. These histograms were obtained by simulating $10^5$ instances of PA decoding for each value of $p$. For each instance, the PA decoder used a high number of resources, specifically $R=1000$ replicas, $N_T=100$ temperature steps, and $N_S=200$ sweeps, to achieve a small error in the histograms. Since the number of incorrect decodings for $p=0.04$ is small, we include an inset with a zoom in the relevant region.}
\label{fig:histograms_p}
\end{figure}

Using this approach, it is possible to set a maximum target value of $\Delta p_L$ and find the corresponding value of $\text{Var}(\beta\Delta F)$. The optimization problem is then transformed into finding the value of the parameters (number of replicas, number of temperature steps, and number of spin flips per temperature) that achieves that variance value. Although one can simplify this three-dimensional problem by fixing two of these parameters while scanning different values of the remaining one (similar to what is shown in Fig.~\ref{fig:error_estimation_benchmark}), a better optimization would require a three-dimensional scan of the parameters.

\begin{figure}[ht]
\centering
\includegraphics[width=0.99\columnwidth]{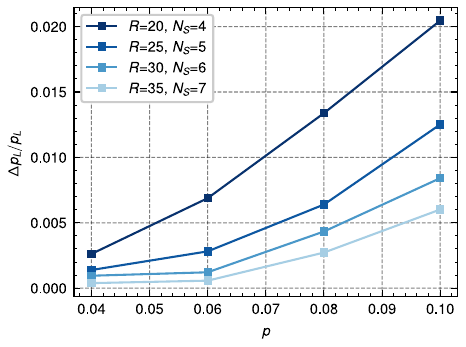}
\caption{Behaviour of the relative error, $\Delta p_L/p_L$, due to finite resources for a code distance $d=11$ and for different values of $p$ and values of $R N_S$ chosen to be $N_S=4,5,6,7$ sweeps and $R=20, 25, 30, 35$ replicas, with $N_T=100$ temperature steps. These results are obtained using histograms such as those shown in Fig.~\ref{fig:histograms_p}. As can be seen, given a number of computational resources, $R N_S$, the relative error decreases with $p$. This means that for a target relative error, $\Delta p_L/p_L$, the number of computational resources can be decreased as $p$ decreases.}
\label{fig:resources_below_threshold}
\end{figure}

We conclude this section by noting that a similar analysis can be performed by fixing the distance of the code and considering different values of $p$. An example of this process is shown in Fig.~\ref{fig:histograms_p} for $d=11$. We can see that, given some computational resources $R N_S$, and for decreasing values of $p$ below the threshold, the values of $|\beta \Delta F|$ move away from zero. Thus, it is expected for the value $\Delta p_L$ to decrease given a fixed value of $\text{Var}(\beta\Delta F)$. Moreover, using the histogram analysis, we see that the value $\Delta p_L/p_L$ also decreases with $p$ (see Fig.~\ref{fig:resources_below_threshold}), with the case close to the threshold being the most expensive case in terms of computational resources. This insight can be useful for reducing the decoding times in codes with error probabilities far from the threshold.

\section{Conclusions and outlook}
\label{sec:conclusions}

In this work, we have shown an implementation of the population annealing algorithm as a color code decoder. This algorithm is based on a mapping of the color code lattice to a spin model, as shown in Ref.~\cite{takada2024ising}. We introduce the use of population annealing, allowing the estimation of the free energy of the different homology classes. This can be used to infer the most probable error class instead of the most probable error case. As a result, we obtain improved thresholds and a higher decoding success rate, which leads to lower logical error rates for the same physical error rate. Our numerical results show that our decoder can reach near-optimal thresholds under code capacity noise (bit-flip and depolarizing) and a high threshold for phenomenological noise.

We provide methods to optimize the hyperparameters of the algorithm, thus reducing the computational resources and time required for a given performance. Additionally, considerable efforts have been made to optimize the code that implements the population annealing decoder. However, the computational runtime required could become a limiting factor when scaling the lattice, preventing the possibility of a real-time decoder implementation for fault-tolerant quantum computation. This can be more challenging in platforms with very fast gates like superconducting qubits, where a QEC cycle can be executed in under $1\mu s$. Nevertheless, some changes could further increase the speed of the decoder. From the algorithmic side, the PA algorithm can perform multiple independent runs with fewer replicas each and perform a weighted average of the results. This results in a reduction of the statistical and systematic errors as compared to making a single PA run with all the replicas~\cite{machta2010population, ebert2022weighted}, which is the case for the implementation in this work. Furthermore, studying possible cluster updates applicable to the color code spin model would be interesting. Cluster updates, as opposed to the single-spin flips used by our algorithm, could reduce the steps needed for thermalization~\cite{houdayer2001cluster, zhu2015efficient}. From the hardware side, our decoder, similar to the one shown in Ref.~\cite{takada2024ising}, is highly parallelizable. While one could take advantage of this by using better CPUs with more parallelization capabilities, or even multiple CPUs, we believe that the most interesting approach would be the implementation of the code in GPUs. The PA algorithm has already been implemented in GPUs, achieving impressive improvements in the performance of the algorithm~\cite{weigel2012performance, barash2017gpu}, considerably reducing the average time required per spin flip thanks to the parallelization capabilities of GPUs. We leave the study of the performance of the PA decoder using a GPU implementation and the analysis of the scaling of computational resources for future work. Finally, similar to the SA decoder, a trade-off exists between performance and decoding time. Thus, one could also decrease the decoding quality in exchange for a faster decoding algorithm.

From the QEC perspective, it would be interesting to understand the discrepancy between our threshold and the estimated optimal threshold for the phenomenological noise model. Finally, we note that, while we have focused on the study of the decoder applied to color codes, the algorithm can be easily applied to other stabilizer codes, like surface codes or qLDPC codes~\cite{gottesman2014faulttolerant,kovalev2013qLDPC,breuckmann2021qLDPC,viszlai2023qLDPC,koukoulekidis2024qldpc,bravyi2023qLDPC}. Furthermore, this could be adapted to better represent circuit-level noise models by introducing additional couplings in the spin model related to the possible errors in the system, as has been shown in Ref.~\cite{vodola2022repetition}. These applications are outside of the scope of this study and are left for future work.

\begin{acknowledgments}
The authors thank Martin Leib for useful discussions.
F.M.-G. also gratefully acknowledges the Scientific Computing Area (AIC), SGAI-CSIC, for their assistance while using the DRAGO Supercomputer for performing the simulations, and support from the Spanish project PID2021-127968NB-I00 funded by MICIU/AEI/10.13039/501100011033 and FEDER, UE, and the CSIC Research Platform on Quantum Technologies PTI-001.
\end{acknowledgments}

\appendix

\bibliography{sample}

\begin{thebibliography}{72}%
\makeatletter
\providecommand \@ifxundefined [1]{%
 \@ifx{#1\undefined}
}%
\providecommand \@ifnum [1]{%
 \ifnum #1\expandafter \@firstoftwo
 \else \expandafter \@secondoftwo
 \fi
}%
\providecommand \@ifx [1]{%
 \ifx #1\expandafter \@firstoftwo
 \else \expandafter \@secondoftwo
 \fi
}%
\providecommand \natexlab [1]{#1}%
\providecommand \enquote  [1]{``#1''}%
\providecommand \bibnamefont  [1]{#1}%
\providecommand \bibfnamefont [1]{#1}%
\providecommand \citenamefont [1]{#1}%
\providecommand \href@noop [0]{\@secondoftwo}%
\providecommand \href [0]{\begingroup \@sanitize@url \@href}%
\providecommand \@href[1]{\@@startlink{#1}\@@href}%
\providecommand \@@href[1]{\endgroup#1\@@endlink}%
\providecommand \@sanitize@url [0]{\catcode `\\12\catcode `\$12\catcode
  `\&12\catcode `\#12\catcode `\^12\catcode `\_12\catcode `\%12\relax}%
\providecommand \@@startlink[1]{}%
\providecommand \@@endlink[0]{}%
\providecommand \url  [0]{\begingroup\@sanitize@url \@url }%
\providecommand \@url [1]{\endgroup\@href {#1}{\urlprefix }}%
\providecommand \urlprefix  [0]{URL }%
\providecommand \Eprint [0]{\href }%
\providecommand \doibase [0]{https://doi.org/}%
\providecommand \selectlanguage [0]{\@gobble}%
\providecommand \bibinfo  [0]{\@secondoftwo}%
\providecommand \bibfield  [0]{\@secondoftwo}%
\providecommand \translation [1]{[#1]}%
\providecommand \BibitemOpen [0]{}%
\providecommand \bibitemStop [0]{}%
\providecommand \bibitemNoStop [0]{.\EOS\space}%
\providecommand \EOS [0]{\spacefactor3000\relax}%
\providecommand \BibitemShut  [1]{\csname bibitem#1\endcsname}%
\let\auto@bib@innerbib\@empty
\bibitem [{\citenamefont {Clarke}\ and\ \citenamefont
  {Wilhelm}(2008)}]{clarke2008superconducting}%
  \BibitemOpen
  \bibfield  {author} {\bibinfo {author} {\bibfnamefont {J.}~\bibnamefont
  {Clarke}}\ and\ \bibinfo {author} {\bibfnamefont {F.~K.}\ \bibnamefont
  {Wilhelm}},\ }\bibfield  {title} {\bibinfo {title} {Superconducting quantum
  bits},\ }\href {https://doi.org/https://doi.org/10.1038/nature07128}
  {\bibfield  {journal} {\bibinfo  {journal} {Nature}\ }\textbf {\bibinfo
  {volume} {453}},\ \bibinfo {pages} {1031} (\bibinfo {year}
  {2008})}\BibitemShut {NoStop}%
\bibitem [{\citenamefont {Paraoanu}(2014)}]{paraoanu2014qsimSuperconducting}%
  \BibitemOpen
  \bibfield  {author} {\bibinfo {author} {\bibfnamefont {G.~S.}\ \bibnamefont
  {Paraoanu}},\ }\bibfield  {title} {\bibinfo {title} {{Recent Progress in
  Quantum Simulation Using Superconducting Circuits}},\ }\href
  {https://doi.org/10.1007/s10909-014-1175-8} {\bibfield  {journal} {\bibinfo
  {journal} {Journal of Low Temperature Physics}\ }\textbf {\bibinfo {volume}
  {175}},\ \bibinfo {pages} {633} (\bibinfo {year} {2014})}\BibitemShut
  {NoStop}%
\bibitem [{\citenamefont {Wendin}(2017)}]{wendin2017quantum}%
  \BibitemOpen
  \bibfield  {author} {\bibinfo {author} {\bibfnamefont {G.}~\bibnamefont
  {Wendin}},\ }\bibfield  {title} {\bibinfo {title} {Quantum information
  processing with superconducting circuits: a review},\ }\href
  {https://doi.org/10.1088/1361-6633/aa7e1a} {\bibfield  {journal} {\bibinfo
  {journal} {Reports on Progress in Physics}\ }\textbf {\bibinfo {volume}
  {80}},\ \bibinfo {pages} {106001} (\bibinfo {year} {2017})}\BibitemShut
  {NoStop}%
\bibitem [{\citenamefont {Kjaergaard}\ \emph {et~al.}(2020)\citenamefont
  {Kjaergaard}, \citenamefont {Schwartz}, \citenamefont {Braum{\"u}ller},
  \citenamefont {Krantz}, \citenamefont {Wang}, \citenamefont {Gustavsson},\
  and\ \citenamefont {Oliver}}]{kjaergaard2020superconducting}%
  \BibitemOpen
  \bibfield  {author} {\bibinfo {author} {\bibfnamefont {M.}~\bibnamefont
  {Kjaergaard}}, \bibinfo {author} {\bibfnamefont {M.~E.}\ \bibnamefont
  {Schwartz}}, \bibinfo {author} {\bibfnamefont {J.}~\bibnamefont
  {Braum{\"u}ller}}, \bibinfo {author} {\bibfnamefont {P.}~\bibnamefont
  {Krantz}}, \bibinfo {author} {\bibfnamefont {J.~I.-J.}\ \bibnamefont {Wang}},
  \bibinfo {author} {\bibfnamefont {S.}~\bibnamefont {Gustavsson}},\ and\
  \bibinfo {author} {\bibfnamefont {W.~D.}\ \bibnamefont {Oliver}},\ }\bibfield
   {title} {\bibinfo {title} {{Superconducting qubits: Current state of
  play}},\ }\href
  {https://doi.org/https://doi.org/10.1146/annurev-conmatphys-031119-050605}
  {\bibfield  {journal} {\bibinfo  {journal} {Annual Review of Condensed Matter
  Physics}\ }\textbf {\bibinfo {volume} {11}},\ \bibinfo {pages} {369}
  (\bibinfo {year} {2020})}\BibitemShut {NoStop}%
\bibitem [{\citenamefont {Cirac}\ and\ \citenamefont
  {Zoller}(1995)}]{cirac1995quantum}%
  \BibitemOpen
  \bibfield  {author} {\bibinfo {author} {\bibfnamefont {J.~I.}\ \bibnamefont
  {Cirac}}\ and\ \bibinfo {author} {\bibfnamefont {P.}~\bibnamefont {Zoller}},\
  }\bibfield  {title} {\bibinfo {title} {Quantum computations with cold trapped
  ions},\ }\href {https://doi.org/10.1103/PhysRevLett.74.4091} {\bibfield
  {journal} {\bibinfo  {journal} {Phys. Rev. Lett.}\ }\textbf {\bibinfo
  {volume} {74}},\ \bibinfo {pages} {4091} (\bibinfo {year}
  {1995})}\BibitemShut {NoStop}%
\bibitem [{\citenamefont {H{\"a}ffner}\ \emph {et~al.}(2008)\citenamefont
  {H{\"a}ffner}, \citenamefont {Roos},\ and\ \citenamefont
  {Blatt}}]{haffner2008quantum}%
  \BibitemOpen
  \bibfield  {author} {\bibinfo {author} {\bibfnamefont {H.}~\bibnamefont
  {H{\"a}ffner}}, \bibinfo {author} {\bibfnamefont {C.~F.}\ \bibnamefont
  {Roos}},\ and\ \bibinfo {author} {\bibfnamefont {R.}~\bibnamefont {Blatt}},\
  }\bibfield  {title} {\bibinfo {title} {Quantum computing with trapped ions},\
  }\href {https://doi.org/https://doi.org/10.1016/j.physrep.2008.09.003}
  {\bibfield  {journal} {\bibinfo  {journal} {Physics reports}\ }\textbf
  {\bibinfo {volume} {469}},\ \bibinfo {pages} {155} (\bibinfo {year}
  {2008})}\BibitemShut {NoStop}%
\bibitem [{\citenamefont {Barreiro}\ \emph {et~al.}(2011)\citenamefont
  {Barreiro}, \citenamefont {M\"{u}ller}, \citenamefont {Schindler},
  \citenamefont {Nigg}, \citenamefont {Monz}, \citenamefont {Chwalla},
  \citenamefont {Hennrich}, \citenamefont {Roos}, \citenamefont {Zoller},\ and\
  \citenamefont {Blatt}}]{barreiro-nature-470-486}%
  \BibitemOpen
  \bibfield  {author} {\bibinfo {author} {\bibfnamefont {J.}~\bibnamefont
  {Barreiro}}, \bibinfo {author} {\bibfnamefont {M.}~\bibnamefont
  {M\"{u}ller}}, \bibinfo {author} {\bibfnamefont {P.}~\bibnamefont
  {Schindler}}, \bibinfo {author} {\bibfnamefont {D.}~\bibnamefont {Nigg}},
  \bibinfo {author} {\bibfnamefont {T.}~\bibnamefont {Monz}}, \bibinfo {author}
  {\bibfnamefont {M.}~\bibnamefont {Chwalla}}, \bibinfo {author} {\bibfnamefont
  {M.}~\bibnamefont {Hennrich}}, \bibinfo {author} {\bibfnamefont {C.~F.}\
  \bibnamefont {Roos}}, \bibinfo {author} {\bibfnamefont {P.}~\bibnamefont
  {Zoller}},\ and\ \bibinfo {author} {\bibfnamefont {R.}~\bibnamefont
  {Blatt}},\ }\bibfield  {title} {\bibinfo {title} {An open-system quantum
  simulator with trapped ions},\ }\href {https://doi.org/10.1038/nature09801}
  {\bibfield  {journal} {\bibinfo  {journal} {Nature}\ }\textbf {\bibinfo
  {volume} {470}},\ \bibinfo {pages} {486} (\bibinfo {year}
  {2011})}\BibitemShut {NoStop}%
\bibitem [{\citenamefont {Lanyon}\ \emph {et~al.}(2011)\citenamefont {Lanyon},
  \citenamefont {Hempel}, \citenamefont {Nigg}, \citenamefont {Müller},
  \citenamefont {Gerritsma}, \citenamefont {Zähringer}, \citenamefont
  {Schindler}, \citenamefont {Barreiro}, \citenamefont {Rambach}, \citenamefont
  {Kirchmair}, \citenamefont {Hennrich}, \citenamefont {Zoller}, \citenamefont
  {Blatt},\ and\ \citenamefont {Roos}}]{lanyon2011digitalQSim}%
  \BibitemOpen
  \bibfield  {author} {\bibinfo {author} {\bibfnamefont {B.~P.}\ \bibnamefont
  {Lanyon}}, \bibinfo {author} {\bibfnamefont {C.}~\bibnamefont {Hempel}},
  \bibinfo {author} {\bibfnamefont {D.}~\bibnamefont {Nigg}}, \bibinfo {author}
  {\bibfnamefont {M.}~\bibnamefont {Müller}}, \bibinfo {author} {\bibfnamefont
  {R.}~\bibnamefont {Gerritsma}}, \bibinfo {author} {\bibfnamefont
  {F.}~\bibnamefont {Zähringer}}, \bibinfo {author} {\bibfnamefont
  {P.}~\bibnamefont {Schindler}}, \bibinfo {author} {\bibfnamefont {J.~T.}\
  \bibnamefont {Barreiro}}, \bibinfo {author} {\bibfnamefont {M.}~\bibnamefont
  {Rambach}}, \bibinfo {author} {\bibfnamefont {G.}~\bibnamefont {Kirchmair}},
  \bibinfo {author} {\bibfnamefont {M.}~\bibnamefont {Hennrich}}, \bibinfo
  {author} {\bibfnamefont {P.}~\bibnamefont {Zoller}}, \bibinfo {author}
  {\bibfnamefont {R.}~\bibnamefont {Blatt}},\ and\ \bibinfo {author}
  {\bibfnamefont {C.~F.}\ \bibnamefont {Roos}},\ }\bibfield  {title} {\bibinfo
  {title} {{Universal Digital Quantum Simulation with Trapped Ions}},\ }\href
  {https://doi.org/10.1126/science.1208001} {\bibfield  {journal} {\bibinfo
  {journal} {Science}\ }\textbf {\bibinfo {volume} {334}},\ \bibinfo {pages}
  {57} (\bibinfo {year} {2011})}\BibitemShut {NoStop}%
\bibitem [{\citenamefont {Brown}\ \emph {et~al.}(2016)\citenamefont {Brown},
  \citenamefont {Kim},\ and\ \citenamefont {Monroe}}]{brown2016co}%
  \BibitemOpen
  \bibfield  {author} {\bibinfo {author} {\bibfnamefont {K.~R.}\ \bibnamefont
  {Brown}}, \bibinfo {author} {\bibfnamefont {J.}~\bibnamefont {Kim}},\ and\
  \bibinfo {author} {\bibfnamefont {C.}~\bibnamefont {Monroe}},\ }\bibfield
  {title} {\bibinfo {title} {Co-designing a scalable quantum computer with
  trapped atomic ions},\ }\href
  {https://doi.org/https://doi.org/10.1038/npjqi.2016.34} {\bibfield  {journal}
  {\bibinfo  {journal} {npj Quantum Information}\ }\textbf {\bibinfo {volume}
  {2}},\ \bibinfo {pages} {1} (\bibinfo {year} {2016})}\BibitemShut {NoStop}%
\bibitem [{\citenamefont {Bruzewicz}\ \emph {et~al.}(2019)\citenamefont
  {Bruzewicz}, \citenamefont {Chiaverini}, \citenamefont {McConnell},\ and\
  \citenamefont {Sage}}]{bruzewicz2019trapped}%
  \BibitemOpen
  \bibfield  {author} {\bibinfo {author} {\bibfnamefont {C.~D.}\ \bibnamefont
  {Bruzewicz}}, \bibinfo {author} {\bibfnamefont {J.}~\bibnamefont
  {Chiaverini}}, \bibinfo {author} {\bibfnamefont {R.}~\bibnamefont
  {McConnell}},\ and\ \bibinfo {author} {\bibfnamefont {J.~M.}\ \bibnamefont
  {Sage}},\ }\bibfield  {title} {\bibinfo {title} {Trapped-ion quantum
  computing: Progress and challenges},\ }\href
  {https://doi.org/https://doi.org/10.1063/1.5088164} {\bibfield  {journal}
  {\bibinfo  {journal} {Applied Physics Reviews}\ }\textbf {\bibinfo {volume}
  {6}},\ \bibinfo {pages} {021314} (\bibinfo {year} {2019})}\BibitemShut
  {NoStop}%
\bibitem [{\citenamefont {Jaksch}\ \emph {et~al.}(2000)\citenamefont {Jaksch},
  \citenamefont {Cirac}, \citenamefont {Zoller}, \citenamefont {Rolston},
  \citenamefont {C{\^o}t{\'e}},\ and\ \citenamefont {Lukin}}]{jaksch2000fast}%
  \BibitemOpen
  \bibfield  {author} {\bibinfo {author} {\bibfnamefont {D.}~\bibnamefont
  {Jaksch}}, \bibinfo {author} {\bibfnamefont {J.~I.}\ \bibnamefont {Cirac}},
  \bibinfo {author} {\bibfnamefont {P.}~\bibnamefont {Zoller}}, \bibinfo
  {author} {\bibfnamefont {S.~L.}\ \bibnamefont {Rolston}}, \bibinfo {author}
  {\bibfnamefont {R.}~\bibnamefont {C{\^o}t{\'e}}},\ and\ \bibinfo {author}
  {\bibfnamefont {M.~D.}\ \bibnamefont {Lukin}},\ }\bibfield  {title} {\bibinfo
  {title} {Fast quantum gates for neutral atoms},\ }\href
  {https://doi.org/10.1103/PhysRevLett.85.2208} {\bibfield  {journal} {\bibinfo
   {journal} {Phys. Rev. Lett.}\ }\textbf {\bibinfo {volume} {85}},\ \bibinfo
  {pages} {2208} (\bibinfo {year} {2000})}\BibitemShut {NoStop}%
\bibitem [{\citenamefont {Saffman}\ \emph {et~al.}(2010)\citenamefont
  {Saffman}, \citenamefont {Walker},\ and\ \citenamefont
  {M{\o}lmer}}]{saffman2010quantum}%
  \BibitemOpen
  \bibfield  {author} {\bibinfo {author} {\bibfnamefont {M.}~\bibnamefont
  {Saffman}}, \bibinfo {author} {\bibfnamefont {T.~G.}\ \bibnamefont
  {Walker}},\ and\ \bibinfo {author} {\bibfnamefont {K.}~\bibnamefont
  {M{\o}lmer}},\ }\bibfield  {title} {\bibinfo {title} {{Quantum information
  with Rydberg atoms}},\ }\href {https://doi.org/10.1103/RevModPhys.82.2313}
  {\bibfield  {journal} {\bibinfo  {journal} {Rev. Mod. Phys.}\ }\textbf
  {\bibinfo {volume} {82}},\ \bibinfo {pages} {2313} (\bibinfo {year}
  {2010})}\BibitemShut {NoStop}%
\bibitem [{\citenamefont {Anderson}\ \emph {et~al.}(2011)\citenamefont
  {Anderson}, \citenamefont {Younge},\ and\ \citenamefont
  {Raithel}}]{anderson2011trapping}%
  \BibitemOpen
  \bibfield  {author} {\bibinfo {author} {\bibfnamefont {S.~E.}\ \bibnamefont
  {Anderson}}, \bibinfo {author} {\bibfnamefont {K.}~\bibnamefont {Younge}},\
  and\ \bibinfo {author} {\bibfnamefont {G.}~\bibnamefont {Raithel}},\
  }\bibfield  {title} {\bibinfo {title} {{Trapping Rydberg atoms in an optical
  lattice}},\ }\href {https://doi.org/10.1103/PhysRevLett.107.263001}
  {\bibfield  {journal} {\bibinfo  {journal} {Phys. Rev. Lett.}\ }\textbf
  {\bibinfo {volume} {107}},\ \bibinfo {pages} {263001} (\bibinfo {year}
  {2011})}\BibitemShut {NoStop}%
\bibitem [{\citenamefont {Bluvstein}\ \emph {et~al.}(2022)\citenamefont
  {Bluvstein}, \citenamefont {Levine}, \citenamefont {Semeghini}, \citenamefont
  {Wang}, \citenamefont {Ebadi}, \citenamefont {Kalinowski}, \citenamefont
  {Keesling}, \citenamefont {Maskara}, \citenamefont {Pichler}, \citenamefont
  {Greiner} \emph {et~al.}}]{bluvstein2022quantum}%
  \BibitemOpen
  \bibfield  {author} {\bibinfo {author} {\bibfnamefont {D.}~\bibnamefont
  {Bluvstein}}, \bibinfo {author} {\bibfnamefont {H.}~\bibnamefont {Levine}},
  \bibinfo {author} {\bibfnamefont {G.}~\bibnamefont {Semeghini}}, \bibinfo
  {author} {\bibfnamefont {T.~T.}\ \bibnamefont {Wang}}, \bibinfo {author}
  {\bibfnamefont {S.}~\bibnamefont {Ebadi}}, \bibinfo {author} {\bibfnamefont
  {M.}~\bibnamefont {Kalinowski}}, \bibinfo {author} {\bibfnamefont
  {A.}~\bibnamefont {Keesling}}, \bibinfo {author} {\bibfnamefont
  {N.}~\bibnamefont {Maskara}}, \bibinfo {author} {\bibfnamefont
  {H.}~\bibnamefont {Pichler}}, \bibinfo {author} {\bibfnamefont
  {M.}~\bibnamefont {Greiner}}, \emph {et~al.},\ }\bibfield  {title} {\bibinfo
  {title} {A quantum processor based on coherent transport of entangled atom
  arrays},\ }\href {https://doi.org/https://doi.org/10.1038/s41586-022-04592-6}
  {\bibfield  {journal} {\bibinfo  {journal} {Nature}\ }\textbf {\bibinfo
  {volume} {604}},\ \bibinfo {pages} {451} (\bibinfo {year}
  {2022})}\BibitemShut {NoStop}%
\bibitem [{\citenamefont {Bluvstein}\ \emph {et~al.}(2024)\citenamefont
  {Bluvstein}, \citenamefont {Evered}, \citenamefont {Geim}, \citenamefont
  {Li}, \citenamefont {Zhou}, \citenamefont {Manovitz}, \citenamefont {Ebadi},
  \citenamefont {Cain}, \citenamefont {Kalinowski}, \citenamefont {Hangleiter}
  \emph {et~al.}}]{bluvstein2024logical}%
  \BibitemOpen
  \bibfield  {author} {\bibinfo {author} {\bibfnamefont {D.}~\bibnamefont
  {Bluvstein}}, \bibinfo {author} {\bibfnamefont {S.~J.}\ \bibnamefont
  {Evered}}, \bibinfo {author} {\bibfnamefont {A.~A.}\ \bibnamefont {Geim}},
  \bibinfo {author} {\bibfnamefont {S.~H.}\ \bibnamefont {Li}}, \bibinfo
  {author} {\bibfnamefont {H.}~\bibnamefont {Zhou}}, \bibinfo {author}
  {\bibfnamefont {T.}~\bibnamefont {Manovitz}}, \bibinfo {author}
  {\bibfnamefont {S.}~\bibnamefont {Ebadi}}, \bibinfo {author} {\bibfnamefont
  {M.}~\bibnamefont {Cain}}, \bibinfo {author} {\bibfnamefont {M.}~\bibnamefont
  {Kalinowski}}, \bibinfo {author} {\bibfnamefont {D.}~\bibnamefont
  {Hangleiter}}, \emph {et~al.},\ }\bibfield  {title} {\bibinfo {title}
  {Logical quantum processor based on reconfigurable atom arrays},\ }\href
  {https://doi.org/https://doi.org/10.1038/s41586-023-06927-3} {\bibfield
  {journal} {\bibinfo  {journal} {Nature}\ }\textbf {\bibinfo {volume} {626}},\
  \bibinfo {pages} {58} (\bibinfo {year} {2024})}\BibitemShut {NoStop}%
\bibitem [{\citenamefont {Barz}(2015)}]{barz2015quantum}%
  \BibitemOpen
  \bibfield  {author} {\bibinfo {author} {\bibfnamefont {S.}~\bibnamefont
  {Barz}},\ }\bibfield  {title} {\bibinfo {title} {Quantum computing with
  photons: introduction to the circuit model, the one-way quantum computer, and
  the fundamental principles of photonic experiments},\ }\href
  {https://doi.org/10.1088/0953-4075/48/8/083001} {\bibfield  {journal}
  {\bibinfo  {journal} {Journal of Physics B: Atomic, Molecular and Optical
  Physics}\ }\textbf {\bibinfo {volume} {48}},\ \bibinfo {pages} {083001}
  (\bibinfo {year} {2015})}\BibitemShut {NoStop}%
\bibitem [{\citenamefont {Nielsen}\ and\ \citenamefont
  {Chuang}(2000)}]{nielsen-book}%
  \BibitemOpen
  \bibfield  {author} {\bibinfo {author} {\bibfnamefont {M.~A.}\ \bibnamefont
  {Nielsen}}\ and\ \bibinfo {author} {\bibfnamefont {I.~L.}\ \bibnamefont
  {Chuang}},\ }\href {https://doi.org/10.1017/CBO9780511976667} {\emph
  {\bibinfo {title} {{Quantum Computation and Quantum Information}}}}\
  (\bibinfo  {publisher} {Cambridge University Press},\ \bibinfo {year}
  {2000})\BibitemShut {NoStop}%
\bibitem [{\citenamefont {Almudever}\ \emph {et~al.}(2017)\citenamefont
  {Almudever}, \citenamefont {Lao}, \citenamefont {Fu}, \citenamefont
  {Khammassi}, \citenamefont {Ashraf}, \citenamefont {Iorga}, \citenamefont
  {Varsamopoulos}, \citenamefont {Eichler}, \citenamefont {Wallraff},
  \citenamefont {Geck}, \citenamefont {Kruth}, \citenamefont {Knoch},
  \citenamefont {Bluhm},\ and\ \citenamefont
  {Bertels}}]{almudever2017challenges}%
  \BibitemOpen
  \bibfield  {author} {\bibinfo {author} {\bibfnamefont {C.~G.}\ \bibnamefont
  {Almudever}}, \bibinfo {author} {\bibfnamefont {L.}~\bibnamefont {Lao}},
  \bibinfo {author} {\bibfnamefont {X.}~\bibnamefont {Fu}}, \bibinfo {author}
  {\bibfnamefont {N.}~\bibnamefont {Khammassi}}, \bibinfo {author}
  {\bibfnamefont {I.}~\bibnamefont {Ashraf}}, \bibinfo {author} {\bibfnamefont
  {D.}~\bibnamefont {Iorga}}, \bibinfo {author} {\bibfnamefont
  {S.}~\bibnamefont {Varsamopoulos}}, \bibinfo {author} {\bibfnamefont
  {C.}~\bibnamefont {Eichler}}, \bibinfo {author} {\bibfnamefont
  {A.}~\bibnamefont {Wallraff}}, \bibinfo {author} {\bibfnamefont
  {L.}~\bibnamefont {Geck}}, \bibinfo {author} {\bibfnamefont {A.}~\bibnamefont
  {Kruth}}, \bibinfo {author} {\bibfnamefont {J.}~\bibnamefont {Knoch}},
  \bibinfo {author} {\bibfnamefont {H.}~\bibnamefont {Bluhm}},\ and\ \bibinfo
  {author} {\bibfnamefont {K.}~\bibnamefont {Bertels}},\ }\bibfield  {title}
  {\bibinfo {title} {The engineering challenges in quantum computing},\ }in\
  \href {https://doi.org/10.23919/DATE.2017.7927104} {\emph {\bibinfo
  {booktitle} {Design, Automation Test in Europe Conference Exhibition (DATE),
  2017}}}\ (\bibinfo {year} {2017})\ p.\ \bibinfo {pages} {836}\BibitemShut
  {NoStop}%
\bibitem [{\citenamefont {Ratcliff}\ \emph {et~al.}(2017)\citenamefont
  {Ratcliff}, \citenamefont {Mohr}, \citenamefont {Huhs}, \citenamefont
  {Deutsch}, \citenamefont {Masella},\ and\ \citenamefont
  {Genovese}}]{ratcliff2017quantumchallenges}%
  \BibitemOpen
  \bibfield  {author} {\bibinfo {author} {\bibfnamefont {L.~E.}\ \bibnamefont
  {Ratcliff}}, \bibinfo {author} {\bibfnamefont {S.}~\bibnamefont {Mohr}},
  \bibinfo {author} {\bibfnamefont {G.}~\bibnamefont {Huhs}}, \bibinfo {author}
  {\bibfnamefont {T.}~\bibnamefont {Deutsch}}, \bibinfo {author} {\bibfnamefont
  {M.}~\bibnamefont {Masella}},\ and\ \bibinfo {author} {\bibfnamefont
  {L.}~\bibnamefont {Genovese}},\ }\bibfield  {title} {\bibinfo {title}
  {{Challenges in large scale quantum mechanical calculations}},\ }\href
  {https://doi.org/https://doi.org/10.1002/wcms.1290} {\bibfield  {journal}
  {\bibinfo  {journal} {WIREs Computational Molecular Science}\ }\textbf
  {\bibinfo {volume} {7}},\ \bibinfo {pages} {e1290} (\bibinfo {year}
  {2017})}\BibitemShut {NoStop}%
\bibitem [{\citenamefont {Terhal}(2015)}]{Terhal2015}%
  \BibitemOpen
  \bibfield  {author} {\bibinfo {author} {\bibfnamefont {B.~M.}\ \bibnamefont
  {Terhal}},\ }\bibfield  {title} {\bibinfo {title} {Quantum error correction
  for quantum memories},\ }\href {https://doi.org/10.1103/RevModPhys.87.307}
  {\bibfield  {journal} {\bibinfo  {journal} {Rev. Mod. Phys.}\ }\textbf
  {\bibinfo {volume} {87}},\ \bibinfo {pages} {307} (\bibinfo {year}
  {2015})}\BibitemShut {NoStop}%
\bibitem [{\citenamefont {Gottesman}(1997)}]{gottesman1997stabilizer}%
  \BibitemOpen
  \bibfield  {author} {\bibinfo {author} {\bibfnamefont {D.}~\bibnamefont
  {Gottesman}},\ }\href
  {https://doi.org/https://doi.org/10.48550/arXiv.quant-ph/9705052} {\emph
  {\bibinfo {title} {Stabilizer codes and quantum error correction}}}\
  (\bibinfo  {publisher} {California Institute of Technology},\ \bibinfo {year}
  {1997})\BibitemShut {NoStop}%
\bibitem [{\citenamefont {Kitaev}(1997)}]{Kitaev1997}%
  \BibitemOpen
  \bibfield  {author} {\bibinfo {author} {\bibfnamefont {A.~{\relax Yu}.}\
  \bibnamefont {Kitaev}},\ }\bibfield  {title} {\bibinfo {title} {{Quantum
  Error Correction with Imperfect Gates}},\ }in\ \href
  {https://doi.org/10.1007/978-1-4615-5923-8_19} {\emph {\bibinfo {booktitle}
  {{Quantum Communication, Computing, and Measurement}}}}\ (\bibinfo
  {publisher} {Springer, Boston, MA},\ \bibinfo {year} {1997})\ p.\ \bibinfo
  {pages} {181}\BibitemShut {NoStop}%
\bibitem [{\citenamefont {Kitaev}(2003)}]{Kitaev2003}%
  \BibitemOpen
  \bibfield  {author} {\bibinfo {author} {\bibfnamefont {A.}~\bibnamefont
  {Kitaev}},\ }\bibfield  {title} {\bibinfo {title} {Fault-tolerant quantum
  computation by anyons},\ }\href
  {https://doi.org/10.1016/S0003-4916(02)00018-0} {\bibfield  {journal}
  {\bibinfo  {journal} {Ann. Phys.}\ }\textbf {\bibinfo {volume} {303}},\
  \bibinfo {pages} {2} (\bibinfo {year} {2003})}\BibitemShut {NoStop}%
\bibitem [{\citenamefont {Dennis}\ \emph {et~al.}(2002)\citenamefont {Dennis},
  \citenamefont {Kitaev}, \citenamefont {Landahl},\ and\ \citenamefont
  {Preskill}}]{dennis2002topological}%
  \BibitemOpen
  \bibfield  {author} {\bibinfo {author} {\bibfnamefont {E.}~\bibnamefont
  {Dennis}}, \bibinfo {author} {\bibfnamefont {A.}~\bibnamefont {Kitaev}},
  \bibinfo {author} {\bibfnamefont {A.}~\bibnamefont {Landahl}},\ and\ \bibinfo
  {author} {\bibfnamefont {J.}~\bibnamefont {Preskill}},\ }\bibfield  {title}
  {\bibinfo {title} {Topological quantum memory},\ }\href
  {https://doi.org/10.1063/1.1499754} {\bibfield  {journal} {\bibinfo
  {journal} {Journal of Mathematical Physics}\ }\textbf {\bibinfo {volume}
  {43}},\ \bibinfo {pages} {4452} (\bibinfo {year} {2002})}\BibitemShut
  {NoStop}%
\bibitem [{\citenamefont {Bombin}\ and\ \citenamefont
  {Martin-Delgado}(2006)}]{colorCodes2006bombin}%
  \BibitemOpen
  \bibfield  {author} {\bibinfo {author} {\bibfnamefont {H.}~\bibnamefont
  {Bombin}}\ and\ \bibinfo {author} {\bibfnamefont {M.~A.}\ \bibnamefont
  {Martin-Delgado}},\ }\bibfield  {title} {\bibinfo {title} {Topological
  quantum distillation},\ }\href
  {https://doi.org/10.1103/PhysRevLett.97.180501} {\bibfield  {journal}
  {\bibinfo  {journal} {Physical Review Letters}\ }\textbf {\bibinfo {volume}
  {97}},\ \bibinfo {pages} {180501} (\bibinfo {year} {2006})}\BibitemShut
  {NoStop}%
\bibitem [{\citenamefont {Bombin}\ and\ \citenamefont
  {Martin-Delgado}(2007)}]{bombin2007colorcodes3d}%
  \BibitemOpen
  \bibfield  {author} {\bibinfo {author} {\bibfnamefont {H.}~\bibnamefont
  {Bombin}}\ and\ \bibinfo {author} {\bibfnamefont {M.~A.}\ \bibnamefont
  {Martin-Delgado}},\ }\bibfield  {title} {\bibinfo {title} {{Topological
  Computation without Braiding}},\ }\href
  {https://doi.org/10.1103/physrevlett.98.160502} {\bibfield  {journal}
  {\bibinfo  {journal} {Physical Review Letters}\ }\textbf {\bibinfo {volume}
  {98}},\ \bibinfo {pages} {160502} (\bibinfo {year} {2007})}\BibitemShut
  {NoStop}%
\bibitem [{\citenamefont {Gottesman}(2014)}]{gottesman2014faulttolerant}%
  \BibitemOpen
  \bibfield  {author} {\bibinfo {author} {\bibfnamefont {D.}~\bibnamefont
  {Gottesman}},\ }\href@noop {} {\bibinfo {title} {{Fault-Tolerant Quantum
  Computation with Constant Overhead}}} (\bibinfo {year} {2014}),\ \Eprint
  {https://arxiv.org/abs/1310.2984} {arXiv:1310.2984 [quant-ph]} \BibitemShut
  {NoStop}%
\bibitem [{\citenamefont {Kovalev}\ and\ \citenamefont
  {Pryadko}(2013)}]{kovalev2013qLDPC}%
  \BibitemOpen
  \bibfield  {author} {\bibinfo {author} {\bibfnamefont {A.~A.}\ \bibnamefont
  {Kovalev}}\ and\ \bibinfo {author} {\bibfnamefont {L.~P.}\ \bibnamefont
  {Pryadko}},\ }\bibfield  {title} {\bibinfo {title} {Fault tolerance of
  quantum low-density parity check codes with sublinear distance scaling},\
  }\href {https://doi.org/10.1103/PhysRevA.87.020304} {\bibfield  {journal}
  {\bibinfo  {journal} {Phys. Rev. A}\ }\textbf {\bibinfo {volume} {87}},\
  \bibinfo {pages} {020304} (\bibinfo {year} {2013})}\BibitemShut {NoStop}%
\bibitem [{\citenamefont {Breuckmann}\ and\ \citenamefont
  {Eberhardt}(2021)}]{breuckmann2021qLDPC}%
  \BibitemOpen
  \bibfield  {author} {\bibinfo {author} {\bibfnamefont {N.~P.}\ \bibnamefont
  {Breuckmann}}\ and\ \bibinfo {author} {\bibfnamefont {J.~N.}\ \bibnamefont
  {Eberhardt}},\ }\bibfield  {title} {\bibinfo {title} {{Quantum Low-Density
  Parity-Check Codes}},\ }\href {https://doi.org/10.1103/PRXQuantum.2.040101}
  {\bibfield  {journal} {\bibinfo  {journal} {PRX Quantum}\ }\textbf {\bibinfo
  {volume} {2}},\ \bibinfo {pages} {040101} (\bibinfo {year}
  {2021})}\BibitemShut {NoStop}%
\bibitem [{\citenamefont {Viszlai}\ \emph {et~al.}(2023)\citenamefont
  {Viszlai}, \citenamefont {Yang}, \citenamefont {Lin}, \citenamefont {Liu},
  \citenamefont {Nottingham}, \citenamefont {Baker},\ and\ \citenamefont
  {Chong}}]{viszlai2023qLDPC}%
  \BibitemOpen
  \bibfield  {author} {\bibinfo {author} {\bibfnamefont {J.}~\bibnamefont
  {Viszlai}}, \bibinfo {author} {\bibfnamefont {W.}~\bibnamefont {Yang}},
  \bibinfo {author} {\bibfnamefont {S.~F.}\ \bibnamefont {Lin}}, \bibinfo
  {author} {\bibfnamefont {J.}~\bibnamefont {Liu}}, \bibinfo {author}
  {\bibfnamefont {N.}~\bibnamefont {Nottingham}}, \bibinfo {author}
  {\bibfnamefont {J.~M.}\ \bibnamefont {Baker}},\ and\ \bibinfo {author}
  {\bibfnamefont {F.~T.}\ \bibnamefont {Chong}},\ }\href@noop {} {\bibinfo
  {title} {{Matching Generalized-Bicycle Codes to Neutral Atoms for
  Low-Overhead Fault-Tolerance}}} (\bibinfo {year} {2023}),\ \Eprint
  {https://arxiv.org/abs/2311.16980} {arXiv:2311.16980 [quant-ph]} \BibitemShut
  {NoStop}%
\bibitem [{\citenamefont {Koukoulekidis}\ \emph {et~al.}(2024)\citenamefont
  {Koukoulekidis}, \citenamefont {Šimkovic IV}, \citenamefont {Leib},\ and\
  \citenamefont {Pereira}}]{koukoulekidis2024qldpc}%
  \BibitemOpen
  \bibfield  {author} {\bibinfo {author} {\bibfnamefont {N.}~\bibnamefont
  {Koukoulekidis}}, \bibinfo {author} {\bibfnamefont {F.}~\bibnamefont
  {Šimkovic IV}}, \bibinfo {author} {\bibfnamefont {M.}~\bibnamefont {Leib}},\
  and\ \bibinfo {author} {\bibfnamefont {F.~R.~F.}\ \bibnamefont {Pereira}},\
  }\href@noop {} {\bibinfo {title} {{Small Quantum Codes from Algebraic
  Extensions of Generalized Bicycle Codes}}} (\bibinfo {year} {2024}),\ \Eprint
  {https://arxiv.org/abs/2401.07583} {arXiv:2401.07583 [quant-ph]} \BibitemShut
  {NoStop}%
\bibitem [{\citenamefont {Bravyi}\ \emph {et~al.}(2023)\citenamefont {Bravyi},
  \citenamefont {Cross}, \citenamefont {Gambetta}, \citenamefont {Maslov},
  \citenamefont {Rall},\ and\ \citenamefont {Yoder}}]{bravyi2023qLDPC}%
  \BibitemOpen
  \bibfield  {author} {\bibinfo {author} {\bibfnamefont {S.}~\bibnamefont
  {Bravyi}}, \bibinfo {author} {\bibfnamefont {A.~W.}\ \bibnamefont {Cross}},
  \bibinfo {author} {\bibfnamefont {J.~M.}\ \bibnamefont {Gambetta}}, \bibinfo
  {author} {\bibfnamefont {D.}~\bibnamefont {Maslov}}, \bibinfo {author}
  {\bibfnamefont {P.}~\bibnamefont {Rall}},\ and\ \bibinfo {author}
  {\bibfnamefont {T.~J.}\ \bibnamefont {Yoder}},\ }\href@noop {} {\bibinfo
  {title} {High-threshold and low-overhead fault-tolerant quantum memory}}
  (\bibinfo {year} {2023}),\ \Eprint {https://arxiv.org/abs/2308.07915}
  {arXiv:2308.07915 [quant-ph]} \BibitemShut {NoStop}%
\bibitem [{\citenamefont {Aharonov}\ and\ \citenamefont
  {Ben-Or}(2008)}]{Aharonov2008}%
  \BibitemOpen
  \bibfield  {author} {\bibinfo {author} {\bibfnamefont {D.}~\bibnamefont
  {Aharonov}}\ and\ \bibinfo {author} {\bibfnamefont {M.}~\bibnamefont
  {Ben-Or}},\ }\bibfield  {title} {\bibinfo {title} {{Fault-tolerant quantum
  computation with constant error rate}},\ }\href
  {https://doi.org/10.1137/S0097539799359385} {\bibfield  {journal} {\bibinfo
  {journal} {SIAM Journal on Computing}\ }\textbf {\bibinfo {volume} {38}},\
  \bibinfo {pages} {1207} (\bibinfo {year} {2008})}\BibitemShut {NoStop}%
\bibitem [{\citenamefont {Shor}(1996)}]{Shor1996}%
  \BibitemOpen
  \bibfield  {author} {\bibinfo {author} {\bibfnamefont {P.~W.}\ \bibnamefont
  {Shor}},\ }\bibfield  {title} {\bibinfo {title} {{Fault-tolerant quantum
  computation}},\ }in\ \href {https://doi.org/10.1109/SFCS.1996.548464} {\emph
  {\bibinfo {booktitle} {Proceedings of 37th Conference on Foundations of
  Computer Science}}}\ (\bibinfo  {publisher} {IEEE},\ \bibinfo {year} {1996})\
  p.~\bibinfo {pages} {56}\BibitemShut {NoStop}%
\bibitem [{\citenamefont {Preskill}(1998)}]{Preskill1998}%
  \BibitemOpen
  \bibfield  {author} {\bibinfo {author} {\bibfnamefont {J.}~\bibnamefont
  {Preskill}},\ }\bibfield  {title} {\bibinfo {title} {{Reliable quantum
  computers}},\ }\href {https://doi.org/10.1098/rspa.1998.0167} {\bibfield
  {journal} {\bibinfo  {journal} {Proc. R. Soc. Lond. A.}\ }\textbf {\bibinfo
  {volume} {454}},\ \bibinfo {pages} {385} (\bibinfo {year}
  {1998})}\BibitemShut {NoStop}%
\bibitem [{\citenamefont {Katzgraber}\ \emph {et~al.}(2009)\citenamefont
  {Katzgraber}, \citenamefont {Bomb{\'\i}n},\ and\ \citenamefont
  {Martin-Delgado}}]{katzgraber2009error}%
  \BibitemOpen
  \bibfield  {author} {\bibinfo {author} {\bibfnamefont {H.~G.}\ \bibnamefont
  {Katzgraber}}, \bibinfo {author} {\bibfnamefont {H.}~\bibnamefont
  {Bomb{\'\i}n}},\ and\ \bibinfo {author} {\bibfnamefont {M.~A.}\ \bibnamefont
  {Martin-Delgado}},\ }\bibfield  {title} {\bibinfo {title} {{Error threshold
  for color codes and random three-body Ising models}},\ }\href
  {https://doi.org/10.1103/PhysRevLett.103.090501} {\bibfield  {journal}
  {\bibinfo  {journal} {Physical Review Letters}\ }\textbf {\bibinfo {volume}
  {103}},\ \bibinfo {pages} {090501} (\bibinfo {year} {2009})}\BibitemShut
  {NoStop}%
\bibitem [{\citenamefont {Bombin}\ \emph {et~al.}(2012)\citenamefont {Bombin},
  \citenamefont {Andrist}, \citenamefont {Ohzeki}, \citenamefont {Katzgraber},\
  and\ \citenamefont {Martin-Delgado}}]{bombin2012strong}%
  \BibitemOpen
  \bibfield  {author} {\bibinfo {author} {\bibfnamefont {H.}~\bibnamefont
  {Bombin}}, \bibinfo {author} {\bibfnamefont {R.~S.}\ \bibnamefont {Andrist}},
  \bibinfo {author} {\bibfnamefont {M.}~\bibnamefont {Ohzeki}}, \bibinfo
  {author} {\bibfnamefont {H.~G.}\ \bibnamefont {Katzgraber}},\ and\ \bibinfo
  {author} {\bibfnamefont {M.~A.}\ \bibnamefont {Martin-Delgado}},\ }\bibfield
  {title} {\bibinfo {title} {Strong resilience of topological codes to
  depolarization},\ }\href {https://doi.org/10.1103/PhysRevX.2.021004}
  {\bibfield  {journal} {\bibinfo  {journal} {Phys. Rev. X}\ }\textbf {\bibinfo
  {volume} {2}},\ \bibinfo {pages} {021004} (\bibinfo {year}
  {2012})}\BibitemShut {NoStop}%
\bibitem [{\citenamefont {Andrist}\ \emph {et~al.}(2011)\citenamefont
  {Andrist}, \citenamefont {Katzgraber}, \citenamefont {Bombin},\ and\
  \citenamefont {Martin-Delgado}}]{katzgraber2011tricolored}%
  \BibitemOpen
  \bibfield  {author} {\bibinfo {author} {\bibfnamefont {R.~S.}\ \bibnamefont
  {Andrist}}, \bibinfo {author} {\bibfnamefont {H.~G.}\ \bibnamefont
  {Katzgraber}}, \bibinfo {author} {\bibfnamefont {H.}~\bibnamefont {Bombin}},\
  and\ \bibinfo {author} {\bibfnamefont {M.~A.}\ \bibnamefont
  {Martin-Delgado}},\ }\bibfield  {title} {\bibinfo {title} {Tricolored lattice
  gauge theory with randomness: fault tolerance in topological color codes},\
  }\href {https://doi.org/10.1088/1367-2630/13/8/083006} {\bibfield  {journal}
  {\bibinfo  {journal} {New Journal of Physics}\ }\textbf {\bibinfo {volume}
  {13}},\ \bibinfo {pages} {083006} (\bibinfo {year} {2011})}\BibitemShut
  {NoStop}%
\bibitem [{\citenamefont {Sarvepalli}\ and\ \citenamefont
  {Raussendorf}(2012)}]{sarvepalli2012rescaling}%
  \BibitemOpen
  \bibfield  {author} {\bibinfo {author} {\bibfnamefont {P.}~\bibnamefont
  {Sarvepalli}}\ and\ \bibinfo {author} {\bibfnamefont {R.}~\bibnamefont
  {Raussendorf}},\ }\bibfield  {title} {\bibinfo {title} {Efficient decoding of
  topological color codes},\ }\href
  {https://doi.org/10.1103/PhysRevA.85.022317} {\bibfield  {journal} {\bibinfo
  {journal} {Phys. Rev. A}\ }\textbf {\bibinfo {volume} {85}},\ \bibinfo
  {pages} {022317} (\bibinfo {year} {2012})}\BibitemShut {NoStop}%
\bibitem [{\citenamefont {Wang}\ \emph {et~al.}(2009)\citenamefont {Wang},
  \citenamefont {Fowler}, \citenamefont {Hill},\ and\ \citenamefont
  {Hollenberg}}]{wang2009graphical}%
  \BibitemOpen
  \bibfield  {author} {\bibinfo {author} {\bibfnamefont {D.}~\bibnamefont
  {Wang}}, \bibinfo {author} {\bibfnamefont {A.}~\bibnamefont {Fowler}},
  \bibinfo {author} {\bibfnamefont {C.}~\bibnamefont {Hill}},\ and\ \bibinfo
  {author} {\bibfnamefont {L.}~\bibnamefont {Hollenberg}},\ }\bibfield  {title}
  {\bibinfo {title} {{Graphical algorithms and threshold error rates for the 2D
  color code}},\ }\href {https://dl.acm.org/doi/10.5555/2011464.2011469}
  {\bibfield  {journal} {\bibinfo  {journal} {Quantum Information and
  Computation}\ }\textbf {\bibinfo {volume} {10}},\ \bibinfo {pages} {780}
  (\bibinfo {year} {2009})}\BibitemShut {NoStop}%
\bibitem [{\citenamefont {Stephens}(2014)}]{stephens2014efficient}%
  \BibitemOpen
  \bibfield  {author} {\bibinfo {author} {\bibfnamefont {A.~M.}\ \bibnamefont
  {Stephens}},\ }\href@noop {} {\bibinfo {title} {Efficient fault-tolerant
  decoding of topological color codes}} (\bibinfo {year} {2014}),\ \Eprint
  {https://arxiv.org/abs/1402.3037} {arXiv:1402.3037} \BibitemShut {NoStop}%
\bibitem [{\citenamefont {Maskara}\ \emph {et~al.}(2019)\citenamefont
  {Maskara}, \citenamefont {Kubica},\ and\ \citenamefont
  {Jochym-O’Connor}}]{Maskara2019networks}%
  \BibitemOpen
  \bibfield  {author} {\bibinfo {author} {\bibfnamefont {N.}~\bibnamefont
  {Maskara}}, \bibinfo {author} {\bibfnamefont {A.}~\bibnamefont {Kubica}},\
  and\ \bibinfo {author} {\bibfnamefont {T.}~\bibnamefont
  {Jochym-O’Connor}},\ }\bibfield  {title} {\bibinfo {title} {Advantages of
  versatile neural-network decoding for topological codes},\ }\href
  {https://doi.org/10.1103/physreva.99.052351} {\bibfield  {journal} {\bibinfo
  {journal} {Physical Review A}\ }\textbf {\bibinfo {volume} {99}},\ \bibinfo
  {pages} {052351} (\bibinfo {year} {2019})}\BibitemShut {NoStop}%
\bibitem [{\citenamefont {Delfosse}(2014)}]{Delfosse_2014projection}%
  \BibitemOpen
  \bibfield  {author} {\bibinfo {author} {\bibfnamefont {N.}~\bibnamefont
  {Delfosse}},\ }\bibfield  {title} {\bibinfo {title} {Decoding color codes by
  projection onto surface codes},\ }\href
  {https://doi.org/10.1103/physreva.89.012317} {\bibfield  {journal} {\bibinfo
  {journal} {Physical Review A}\ }\textbf {\bibinfo {volume} {89}},\ \bibinfo
  {pages} {012317} (\bibinfo {year} {2014})}\BibitemShut {NoStop}%
\bibitem [{\citenamefont {Delfosse}\ and\ \citenamefont
  {Nickerson}(2021)}]{delfosse2017almostlinear}%
  \BibitemOpen
  \bibfield  {author} {\bibinfo {author} {\bibfnamefont {N.}~\bibnamefont
  {Delfosse}}\ and\ \bibinfo {author} {\bibfnamefont {N.~H.}\ \bibnamefont
  {Nickerson}},\ }\bibfield  {title} {\bibinfo {title} {Almost-linear time
  decoding algorithm for topological codes},\ }\href
  {https://doi.org/10.22331/q-2021-12-02-595} {\bibfield  {journal} {\bibinfo
  {journal} {Quantum}\ }\textbf {\bibinfo {volume} {5}},\ \bibinfo {pages}
  {595} (\bibinfo {year} {2021})}\BibitemShut {NoStop}%
\bibitem [{\citenamefont {Delfosse}\ and\ \citenamefont
  {Z\'emor}(2020)}]{Delfosse_2020peeling}%
  \BibitemOpen
  \bibfield  {author} {\bibinfo {author} {\bibfnamefont {N.}~\bibnamefont
  {Delfosse}}\ and\ \bibinfo {author} {\bibfnamefont {G.}~\bibnamefont
  {Z\'emor}},\ }\bibfield  {title} {\bibinfo {title} {Linear-time maximum
  likelihood decoding of surface codes over the quantum erasure channel},\
  }\href {https://doi.org/10.1103/PhysRevResearch.2.033042} {\bibfield
  {journal} {\bibinfo  {journal} {Physical Review Research}\ }\textbf {\bibinfo
  {volume} {2}},\ \bibinfo {pages} {033042} (\bibinfo {year}
  {2020})}\BibitemShut {NoStop}%
\bibitem [{\citenamefont {Kubica}\ and\ \citenamefont
  {Preskill}(2019)}]{Kubica2019cellular}%
  \BibitemOpen
  \bibfield  {author} {\bibinfo {author} {\bibfnamefont {A.}~\bibnamefont
  {Kubica}}\ and\ \bibinfo {author} {\bibfnamefont {J.}~\bibnamefont
  {Preskill}},\ }\bibfield  {title} {\bibinfo {title} {{Cellular-Automaton
  Decoders with Provable Thresholds for Topological Codes}},\ }\href
  {https://doi.org/10.1103/physrevlett.123.020501} {\bibfield  {journal}
  {\bibinfo  {journal} {Physical Review Letters}\ }\textbf {\bibinfo {volume}
  {123}},\ \bibinfo {pages} {020501} (\bibinfo {year} {2019})}\BibitemShut
  {NoStop}%
\bibitem [{\citenamefont {Kubica}\ and\ \citenamefont
  {Delfosse}(2023)}]{kubica2019restriction}%
  \BibitemOpen
  \bibfield  {author} {\bibinfo {author} {\bibfnamefont {A.}~\bibnamefont
  {Kubica}}\ and\ \bibinfo {author} {\bibfnamefont {N.}~\bibnamefont
  {Delfosse}},\ }\bibfield  {title} {\bibinfo {title} {Efficient color code
  decoders in $ d\geq 2$ dimensions from toric code decoders},\ }\href
  {https://doi.org/https://doi.org/10.22331/q-2023-02-21-929} {\bibfield
  {journal} {\bibinfo  {journal} {Quantum}\ }\textbf {\bibinfo {volume} {7}},\
  \bibinfo {pages} {929} (\bibinfo {year} {2023})}\BibitemShut {NoStop}%
\bibitem [{\citenamefont {Baireuther}\ \emph {et~al.}(2019)\citenamefont
  {Baireuther}, \citenamefont {Caio}, \citenamefont {Criger}, \citenamefont
  {Beenakker},\ and\ \citenamefont {O’Brien}}]{Baireuther_2019}%
  \BibitemOpen
  \bibfield  {author} {\bibinfo {author} {\bibfnamefont {P.}~\bibnamefont
  {Baireuther}}, \bibinfo {author} {\bibfnamefont {M.~D.}\ \bibnamefont
  {Caio}}, \bibinfo {author} {\bibfnamefont {B.}~\bibnamefont {Criger}},
  \bibinfo {author} {\bibfnamefont {C.~W.~J.}\ \bibnamefont {Beenakker}},\ and\
  \bibinfo {author} {\bibfnamefont {T.~E.}\ \bibnamefont {O’Brien}},\
  }\bibfield  {title} {\bibinfo {title} {Neural network decoder for topological
  color codes with circuit level noise},\ }\href
  {https://doi.org/10.1088/1367-2630/aaf29e} {\bibfield  {journal} {\bibinfo
  {journal} {New Journal of Physics}\ }\textbf {\bibinfo {volume} {21}},\
  \bibinfo {pages} {013003} (\bibinfo {year} {2019})}\BibitemShut {NoStop}%
\bibitem [{\citenamefont {Chamberland}\ and\ \citenamefont
  {Ronagh}(2018)}]{Chamberland_2018}%
  \BibitemOpen
  \bibfield  {author} {\bibinfo {author} {\bibfnamefont {C.}~\bibnamefont
  {Chamberland}}\ and\ \bibinfo {author} {\bibfnamefont {P.}~\bibnamefont
  {Ronagh}},\ }\bibfield  {title} {\bibinfo {title} {Deep neural decoders for
  near term fault-tolerant experiments},\ }\href
  {https://doi.org/10.1088/2058-9565/aad1f7} {\bibfield  {journal} {\bibinfo
  {journal} {Quantum Science and Technology}\ }\textbf {\bibinfo {volume}
  {3}},\ \bibinfo {pages} {044002} (\bibinfo {year} {2018})}\BibitemShut
  {NoStop}%
\bibitem [{\citenamefont {Davaasuren}\ \emph {et~al.}(2020)\citenamefont
  {Davaasuren}, \citenamefont {Suzuki}, \citenamefont {Fujii},\ and\
  \citenamefont {Koashi}}]{Davaasuren_2020}%
  \BibitemOpen
  \bibfield  {author} {\bibinfo {author} {\bibfnamefont {A.}~\bibnamefont
  {Davaasuren}}, \bibinfo {author} {\bibfnamefont {Y.}~\bibnamefont {Suzuki}},
  \bibinfo {author} {\bibfnamefont {K.}~\bibnamefont {Fujii}},\ and\ \bibinfo
  {author} {\bibfnamefont {M.}~\bibnamefont {Koashi}},\ }\bibfield  {title}
  {\bibinfo {title} {General framework for constructing fast and near-optimal
  machine-learning-based decoder of the topological stabilizer codes},\ }\href
  {https://doi.org/10.1103/physrevresearch.2.033399} {\bibfield  {journal}
  {\bibinfo  {journal} {Physical Review Research}\ }\textbf {\bibinfo {volume}
  {2}},\ \bibinfo {pages} {033399} (\bibinfo {year} {2020})}\BibitemShut
  {NoStop}%
\bibitem [{\citenamefont {Chubb}(2021)}]{chubb2021general}%
  \BibitemOpen
  \bibfield  {author} {\bibinfo {author} {\bibfnamefont {C.~T.}\ \bibnamefont
  {Chubb}},\ }\href@noop {} {\bibinfo {title} {{General tensor network decoding
  of 2D Pauli codes}}} (\bibinfo {year} {2021}),\ \Eprint
  {https://arxiv.org/abs/2101.04125} {arXiv:2101.04125} \BibitemShut {NoStop}%
\bibitem [{\citenamefont {Parrado-Rodr\'{\i}guez}\ \emph
  {et~al.}(2022)\citenamefont {Parrado-Rodr\'{\i}guez}, \citenamefont
  {Rispler},\ and\ \citenamefont {M\"uller}}]{parrado2022rescaling}%
  \BibitemOpen
  \bibfield  {author} {\bibinfo {author} {\bibfnamefont {P.}~\bibnamefont
  {Parrado-Rodr\'{\i}guez}}, \bibinfo {author} {\bibfnamefont {M.}~\bibnamefont
  {Rispler}},\ and\ \bibinfo {author} {\bibfnamefont {M.}~\bibnamefont
  {M\"uller}},\ }\bibfield  {title} {\bibinfo {title} {Rescaling decoder for
  two-dimensional topological quantum color codes on 4.8.8 lattices},\ }\href
  {https://doi.org/10.1103/PhysRevA.106.032431} {\bibfield  {journal} {\bibinfo
   {journal} {Phys. Rev. A}\ }\textbf {\bibinfo {volume} {106}},\ \bibinfo
  {pages} {032431} (\bibinfo {year} {2022})}\BibitemShut {NoStop}%
\bibitem [{\citenamefont {deMarti iOlius}\ \emph {et~al.}(2023)\citenamefont
  {deMarti iOlius}, \citenamefont {Fuentes}, \citenamefont {Orús},
  \citenamefont {Crespo},\ and\ \citenamefont {Martinez}}]{iolius2023decoding}%
  \BibitemOpen
  \bibfield  {author} {\bibinfo {author} {\bibfnamefont {A.}~\bibnamefont
  {deMarti iOlius}}, \bibinfo {author} {\bibfnamefont {P.}~\bibnamefont
  {Fuentes}}, \bibinfo {author} {\bibfnamefont {R.}~\bibnamefont {Orús}},
  \bibinfo {author} {\bibfnamefont {P.~M.}\ \bibnamefont {Crespo}},\ and\
  \bibinfo {author} {\bibfnamefont {J.~E.}\ \bibnamefont {Martinez}},\
  }\href@noop {} {\bibinfo {title} {Decoding algorithms for surface codes}}
  (\bibinfo {year} {2023}),\ \Eprint {https://arxiv.org/abs/2307.14989}
  {arXiv:2307.14989 [quant-ph]} \BibitemShut {NoStop}%
\bibitem [{\citenamefont {Berent}\ \emph {et~al.}(2023)\citenamefont {Berent},
  \citenamefont {Burgholzer}, \citenamefont {Derks}, \citenamefont {Eisert},\
  and\ \citenamefont {Wille}}]{berent2023decoding}%
  \BibitemOpen
  \bibfield  {author} {\bibinfo {author} {\bibfnamefont {L.}~\bibnamefont
  {Berent}}, \bibinfo {author} {\bibfnamefont {L.}~\bibnamefont {Burgholzer}},
  \bibinfo {author} {\bibfnamefont {P.-J.~H.}\ \bibnamefont {Derks}}, \bibinfo
  {author} {\bibfnamefont {J.}~\bibnamefont {Eisert}},\ and\ \bibinfo {author}
  {\bibfnamefont {R.}~\bibnamefont {Wille}},\ }\href@noop {} {\bibinfo {title}
  {{Decoding quantum color codes with MaxSAT}}} (\bibinfo {year} {2023}),\
  \Eprint {https://arxiv.org/abs/2303.14237} {arXiv:2303.14237} \BibitemShut
  {NoStop}%
\bibitem [{\citenamefont {Kirkpatrick}\ \emph {et~al.}(1983)\citenamefont
  {Kirkpatrick}, \citenamefont {Gelatt~Jr},\ and\ \citenamefont
  {Vecchi}}]{kirkpatrick1983optimization}%
  \BibitemOpen
  \bibfield  {author} {\bibinfo {author} {\bibfnamefont {S.}~\bibnamefont
  {Kirkpatrick}}, \bibinfo {author} {\bibfnamefont {C.~D.}\ \bibnamefont
  {Gelatt~Jr}},\ and\ \bibinfo {author} {\bibfnamefont {M.~P.}\ \bibnamefont
  {Vecchi}},\ }\bibfield  {title} {\bibinfo {title} {Optimization by simulated
  annealing},\ }\href
  {https://doi.org/https://doi.org/10.1126/science.220.4598.671} {\bibfield
  {journal} {\bibinfo  {journal} {Science}\ }\textbf {\bibinfo {volume}
  {220}},\ \bibinfo {pages} {671} (\bibinfo {year} {1983})}\BibitemShut
  {NoStop}%
\bibitem [{\citenamefont {Takada}\ \emph {et~al.}(2024)\citenamefont {Takada},
  \citenamefont {Takeuchi},\ and\ \citenamefont {Fujii}}]{takada2024ising}%
  \BibitemOpen
  \bibfield  {author} {\bibinfo {author} {\bibfnamefont {Y.}~\bibnamefont
  {Takada}}, \bibinfo {author} {\bibfnamefont {Y.}~\bibnamefont {Takeuchi}},\
  and\ \bibinfo {author} {\bibfnamefont {K.}~\bibnamefont {Fujii}},\ }\bibfield
   {title} {\bibinfo {title} {Ising model formulation for highly accurate
  topological color codes decoding},\ }\href
  {https://doi.org/10.1103/PhysRevResearch.6.013092} {\bibfield  {journal}
  {\bibinfo  {journal} {Physical Review Research}\ }\textbf {\bibinfo {volume}
  {6}},\ \bibinfo {pages} {013092} (\bibinfo {year} {2024})}\BibitemShut
  {NoStop}%
\bibitem [{\citenamefont {Takeuchi}\ \emph {et~al.}(2023)\citenamefont
  {Takeuchi}, \citenamefont {Takada}, \citenamefont {Sakashita}, \citenamefont
  {Fujisaki}, \citenamefont {Oshima}, \citenamefont {Sato},\ and\ \citenamefont
  {Fujii}}]{takeuchi2023comparative}%
  \BibitemOpen
  \bibfield  {author} {\bibinfo {author} {\bibfnamefont {Y.}~\bibnamefont
  {Takeuchi}}, \bibinfo {author} {\bibfnamefont {Y.}~\bibnamefont {Takada}},
  \bibinfo {author} {\bibfnamefont {T.}~\bibnamefont {Sakashita}}, \bibinfo
  {author} {\bibfnamefont {J.}~\bibnamefont {Fujisaki}}, \bibinfo {author}
  {\bibfnamefont {H.}~\bibnamefont {Oshima}}, \bibinfo {author} {\bibfnamefont
  {S.}~\bibnamefont {Sato}},\ and\ \bibinfo {author} {\bibfnamefont
  {K.}~\bibnamefont {Fujii}},\ }\href@noop {} {\bibinfo {title} {Comparative
  study of decoding the surface code using simulated annealing under
  depolarizing noise}} (\bibinfo {year} {2023}),\ \Eprint
  {https://arxiv.org/abs/2311.07973} {arXiv:2311.07973 [quant-ph]} \BibitemShut
  {NoStop}%
\bibitem [{\citenamefont {Ohzeki}(2009{\natexlab{a}})}]{ohzeki2009locations}%
  \BibitemOpen
  \bibfield  {author} {\bibinfo {author} {\bibfnamefont {M.}~\bibnamefont
  {Ohzeki}},\ }\bibfield  {title} {\bibinfo {title} {Locations of multicritical
  points for spin glasses on regular lattices},\ }\href
  {https://doi.org/10.1103/PhysRevE.79.021129} {\bibfield  {journal} {\bibinfo
  {journal} {Physical Review E}\ }\textbf {\bibinfo {volume} {79}},\ \bibinfo
  {pages} {021129} (\bibinfo {year} {2009}{\natexlab{a}})}\BibitemShut
  {NoStop}%
\bibitem [{\citenamefont {Ohzeki}(2009{\natexlab{b}})}]{Ohzeki_2009}%
  \BibitemOpen
  \bibfield  {author} {\bibinfo {author} {\bibfnamefont {M.}~\bibnamefont
  {Ohzeki}},\ }\bibfield  {title} {\bibinfo {title} {Accuracy thresholds of
  topological color codes on the hexagonal and square-octagonal lattices},\
  }\href {https://doi.org/10.1103/physreve.80.011141} {\bibfield  {journal}
  {\bibinfo  {journal} {Physical Review E}\ }\textbf {\bibinfo {volume} {80}},\
  \bibinfo {pages} {011141} (\bibinfo {year} {2009}{\natexlab{b}})}\BibitemShut
  {NoStop}%
\bibitem [{\citenamefont {Hukushima}\ and\ \citenamefont
  {Iba}(2003)}]{hukushima2003population}%
  \BibitemOpen
  \bibfield  {author} {\bibinfo {author} {\bibfnamefont {K.}~\bibnamefont
  {Hukushima}}\ and\ \bibinfo {author} {\bibfnamefont {Y.}~\bibnamefont
  {Iba}},\ }\bibfield  {title} {\bibinfo {title} {Population annealing and its
  application to a spin glass},\ }in\ \href
  {https://doi.org/https://doi.org/10.1063/1.1632130} {\emph {\bibinfo
  {booktitle} {AIP Conference Proceedings}}},\ Vol.\ \bibinfo {volume} {690}\
  (\bibinfo {organization} {American Institute of Physics},\ \bibinfo {year}
  {2003})\ pp.\ \bibinfo {pages} {200--206}\BibitemShut {NoStop}%
\bibitem [{\citenamefont {Machta}(2010)}]{machta2010population}%
  \BibitemOpen
  \bibfield  {author} {\bibinfo {author} {\bibfnamefont {J.}~\bibnamefont
  {Machta}},\ }\bibfield  {title} {\bibinfo {title} {{Population annealing with
  weighted averages: A Monte Carlo method for rough free-energy landscapes}},\
  }\href {https://doi.org/10.1103/PhysRevE.82.026704} {\bibfield  {journal}
  {\bibinfo  {journal} {Physical Review E}\ }\textbf {\bibinfo {volume} {82}},\
  \bibinfo {pages} {026704} (\bibinfo {year} {2010})}\BibitemShut {NoStop}%
\bibitem [{\citenamefont {Ohno}\ \emph {et~al.}(2004)\citenamefont {Ohno},
  \citenamefont {Arakawa}, \citenamefont {Ichinose},\ and\ \citenamefont
  {Matsui}}]{ohno2004phase}%
  \BibitemOpen
  \bibfield  {author} {\bibinfo {author} {\bibfnamefont {T.}~\bibnamefont
  {Ohno}}, \bibinfo {author} {\bibfnamefont {G.}~\bibnamefont {Arakawa}},
  \bibinfo {author} {\bibfnamefont {I.}~\bibnamefont {Ichinose}},\ and\
  \bibinfo {author} {\bibfnamefont {T.}~\bibnamefont {Matsui}},\ }\bibfield
  {title} {\bibinfo {title} {{Phase structure of the random-plaquette Z2 gauge
  model: accuracy threshold for a toric quantum memory}},\ }\href
  {https://doi.org/https://doi.org/10.1016/j.nuclphysb.2004.07.003} {\bibfield
  {journal} {\bibinfo  {journal} {Nuclear physics B}\ }\textbf {\bibinfo
  {volume} {697}},\ \bibinfo {pages} {462} (\bibinfo {year}
  {2004})}\BibitemShut {NoStop}%
\bibitem [{\citenamefont {Andrist}\ \emph {et~al.}(2016)\citenamefont
  {Andrist}, \citenamefont {Katzgraber}, \citenamefont {Bombin},\ and\
  \citenamefont {Martin-Delgado}}]{andrist2016error}%
  \BibitemOpen
  \bibfield  {author} {\bibinfo {author} {\bibfnamefont {R.~S.}\ \bibnamefont
  {Andrist}}, \bibinfo {author} {\bibfnamefont {H.~G.}\ \bibnamefont
  {Katzgraber}}, \bibinfo {author} {\bibfnamefont {H.}~\bibnamefont {Bombin}},\
  and\ \bibinfo {author} {\bibfnamefont {M.}~\bibnamefont {Martin-Delgado}},\
  }\bibfield  {title} {\bibinfo {title} {Error tolerance of topological codes
  with independent bit-flip and measurement errors},\ }\href
  {https://doi.org/10.1103/PhysRevA.94.012318} {\bibfield  {journal} {\bibinfo
  {journal} {Physical Review A}\ }\textbf {\bibinfo {volume} {94}},\ \bibinfo
  {pages} {012318} (\bibinfo {year} {2016})}\BibitemShut {NoStop}%
\bibitem [{\citenamefont {Aaronson}\ and\ \citenamefont
  {Gottesman}(2004)}]{aaronson2004destabilizers}%
  \BibitemOpen
  \bibfield  {author} {\bibinfo {author} {\bibfnamefont {S.}~\bibnamefont
  {Aaronson}}\ and\ \bibinfo {author} {\bibfnamefont {D.}~\bibnamefont
  {Gottesman}},\ }\bibfield  {title} {\bibinfo {title} {Improved simulation of
  stabilizer circuits},\ }\href {https://doi.org/10.1103/physreva.70.052328}
  {\bibfield  {journal} {\bibinfo  {journal} {Physical Review A}\ }\textbf
  {\bibinfo {volume} {70}},\ \bibinfo {pages} {052328} (\bibinfo {year}
  {2004})}\BibitemShut {NoStop}%
\bibitem [{\citenamefont {Robert}\ \emph {et~al.}(1999)\citenamefont {Robert},
  \citenamefont {Casella},\ and\ \citenamefont {Casella}}]{robert1999monte}%
  \BibitemOpen
  \bibfield  {author} {\bibinfo {author} {\bibfnamefont {C.~P.}\ \bibnamefont
  {Robert}}, \bibinfo {author} {\bibfnamefont {G.}~\bibnamefont {Casella}},\
  and\ \bibinfo {author} {\bibfnamefont {G.}~\bibnamefont {Casella}},\ }\href
  {https://doi.org/10.1007/978-1-4757-4145-2} {\emph {\bibinfo {title} {Monte
  Carlo statistical methods}}},\ Vol.~\bibinfo {volume} {2}\ (\bibinfo
  {publisher} {Springer},\ \bibinfo {year} {1999})\BibitemShut {NoStop}%
\bibitem [{\citenamefont {Gessert}\ \emph {et~al.}(2023)\citenamefont
  {Gessert}, \citenamefont {Janke},\ and\ \citenamefont
  {Weigel}}]{gessert2023resampling}%
  \BibitemOpen
  \bibfield  {author} {\bibinfo {author} {\bibfnamefont {D.}~\bibnamefont
  {Gessert}}, \bibinfo {author} {\bibfnamefont {W.}~\bibnamefont {Janke}},\
  and\ \bibinfo {author} {\bibfnamefont {M.}~\bibnamefont {Weigel}},\
  }\bibfield  {title} {\bibinfo {title} {Resampling schemes in population
  annealing: Numerical and theoretical results},\ }\href
  {https://doi.org/10.1103/PhysRevE.108.065309} {\bibfield  {journal} {\bibinfo
   {journal} {Physical Review E}\ }\textbf {\bibinfo {volume} {108}},\ \bibinfo
  {pages} {065309} (\bibinfo {year} {2023})}\BibitemShut {NoStop}%
\bibitem [{\citenamefont {Ebert}\ \emph {et~al.}(2022)\citenamefont {Ebert},
  \citenamefont {Gessert},\ and\ \citenamefont {Weigel}}]{ebert2022weighted}%
  \BibitemOpen
  \bibfield  {author} {\bibinfo {author} {\bibfnamefont {P.~L.}\ \bibnamefont
  {Ebert}}, \bibinfo {author} {\bibfnamefont {D.}~\bibnamefont {Gessert}},\
  and\ \bibinfo {author} {\bibfnamefont {M.}~\bibnamefont {Weigel}},\
  }\bibfield  {title} {\bibinfo {title} {{Weighted averages in population
  annealing: Analysis and general framework}},\ }\href
  {https://doi.org/10.1103/PhysRevE.106.045303} {\bibfield  {journal} {\bibinfo
   {journal} {Physical Review E}\ }\textbf {\bibinfo {volume} {106}},\ \bibinfo
  {pages} {045303} (\bibinfo {year} {2022})}\BibitemShut {NoStop}%
\bibitem [{\citenamefont {Houdayer}(2001)}]{houdayer2001cluster}%
  \BibitemOpen
  \bibfield  {author} {\bibinfo {author} {\bibfnamefont {J.}~\bibnamefont
  {Houdayer}},\ }\bibfield  {title} {\bibinfo {title} {{A cluster Monte Carlo
  algorithm for 2-dimensional spin glasses}},\ }\href
  {https://doi.org/https://doi.org/10.1007/PL00011151} {\bibfield  {journal}
  {\bibinfo  {journal} {The European Physical Journal B-Condensed Matter and
  Complex Systems}\ }\textbf {\bibinfo {volume} {22}},\ \bibinfo {pages} {479}
  (\bibinfo {year} {2001})}\BibitemShut {NoStop}%
\bibitem [{\citenamefont {Zhu}\ \emph {et~al.}(2015)\citenamefont {Zhu},
  \citenamefont {Ochoa},\ and\ \citenamefont {Katzgraber}}]{zhu2015efficient}%
  \BibitemOpen
  \bibfield  {author} {\bibinfo {author} {\bibfnamefont {Z.}~\bibnamefont
  {Zhu}}, \bibinfo {author} {\bibfnamefont {A.~J.}\ \bibnamefont {Ochoa}},\
  and\ \bibinfo {author} {\bibfnamefont {H.~G.}\ \bibnamefont {Katzgraber}},\
  }\bibfield  {title} {\bibinfo {title} {Efficient cluster algorithm for spin
  glasses in any space dimension},\ }\href
  {https://doi.org/10.1103/PhysRevLett.115.077201} {\bibfield  {journal}
  {\bibinfo  {journal} {Physical Review Letters}\ }\textbf {\bibinfo {volume}
  {115}},\ \bibinfo {pages} {077201} (\bibinfo {year} {2015})}\BibitemShut
  {NoStop}%
\bibitem [{\citenamefont {Weigel}(2012)}]{weigel2012performance}%
  \BibitemOpen
  \bibfield  {author} {\bibinfo {author} {\bibfnamefont {M.}~\bibnamefont
  {Weigel}},\ }\bibfield  {title} {\bibinfo {title} {{Performance potential for
  simulating spin models on GPU}},\ }\href
  {https://doi.org/https://doi.org/10.1016/j.jcp.2011.12.008} {\bibfield
  {journal} {\bibinfo  {journal} {Journal of Computational Physics}\ }\textbf
  {\bibinfo {volume} {231}},\ \bibinfo {pages} {3064} (\bibinfo {year}
  {2012})}\BibitemShut {NoStop}%
\bibitem [{\citenamefont {Barash}\ \emph {et~al.}(2017)\citenamefont {Barash},
  \citenamefont {Weigel}, \citenamefont {Borovsk{\`y}}, \citenamefont {Janke},\
  and\ \citenamefont {Shchur}}]{barash2017gpu}%
  \BibitemOpen
  \bibfield  {author} {\bibinfo {author} {\bibfnamefont {L.~Y.}\ \bibnamefont
  {Barash}}, \bibinfo {author} {\bibfnamefont {M.}~\bibnamefont {Weigel}},
  \bibinfo {author} {\bibfnamefont {M.}~\bibnamefont {Borovsk{\`y}}}, \bibinfo
  {author} {\bibfnamefont {W.}~\bibnamefont {Janke}},\ and\ \bibinfo {author}
  {\bibfnamefont {L.~N.}\ \bibnamefont {Shchur}},\ }\bibfield  {title}
  {\bibinfo {title} {{GPU accelerated population annealing algorithm}},\ }\href
  {https://doi.org/https://doi.org/10.1016/j.cpc.2017.06.020} {\bibfield
  {journal} {\bibinfo  {journal} {Computer Physics Communications}\ }\textbf
  {\bibinfo {volume} {220}},\ \bibinfo {pages} {341} (\bibinfo {year}
  {2017})}\BibitemShut {NoStop}%
\bibitem [{\citenamefont {Vodola}\ \emph {et~al.}(2022)\citenamefont {Vodola},
  \citenamefont {Rispler}, \citenamefont {Kim},\ and\ \citenamefont
  {Müller}}]{vodola2022repetition}%
  \BibitemOpen
  \bibfield  {author} {\bibinfo {author} {\bibfnamefont {D.}~\bibnamefont
  {Vodola}}, \bibinfo {author} {\bibfnamefont {M.}~\bibnamefont {Rispler}},
  \bibinfo {author} {\bibfnamefont {S.}~\bibnamefont {Kim}},\ and\ \bibinfo
  {author} {\bibfnamefont {M.}~\bibnamefont {Müller}},\ }\bibfield  {title}
  {\bibinfo {title} {Fundamental thresholds of realistic quantum error
  correction circuits from classical spin models},\ }\href
  {https://doi.org/10.22331/q-2022-01-05-618} {\bibfield  {journal} {\bibinfo
  {journal} {Quantum}\ }\textbf {\bibinfo {volume} {6}},\ \bibinfo {pages}
  {618} (\bibinfo {year} {2022})}\BibitemShut {NoStop}%
\end{thebibliography}%
\end{document}